%% file: BaumannEtAl_arXive20161216.tex
\documentclass[10pt,onecolumn]{article}
\include{Vorspann}

\title{Nonlinear Behavior in High-Intensity Discharge Lamps}

\author{\normalsize Bernd Baumann$^{*1}$,  Joerg Schwieger$^{1,2}$,  Marcus Wolff$^{1}$, \\ \normalsize Freddy Manders$^{3}$, and Jos Suijker$^{3,4}$}

\date{\footnotesize
$^{1}$Hamburg University of Applied Sciences,  Department of Mechanical Engineering and Production,\hfill \phantom{.}\\
\hspace{1mm} Berliner Tor 21, 20099 Hamburg,  Germany\hfill \phantom{.}\\
$^{2}$University of the West of Scotland, School of Engineering and Computing, High Street, \hfill \phantom{.}\\
\hspace{1mm} Paisley PA1 2BE, United Kingdom \hfill\phantom{.}\\
$^{3}$Philips Lighting, Steenweg op Gierle 417, 2300 Turnhout, Belgium \hfill\phantom{.}\\
$^{4}$Technical University Eindhoven, Den Dolech 2, 5612AZ Eindhoven, Netherlands \hfill\phantom{.}\\
$^*$ info@BerndBaumann.de\hfill\phantom{.}}

\begin{document}
\pdfoutput=1
\maketitle

\begin{abstract}
The light flicker problem of high intensity discharge lamps is studied numerically and experimentally. It is shown that in some respects the systems  behaves very similar to the forced Duffing oscillator with a softening spring. In particular, the jump phenomenon and hysteresis are observed in the simulations and in the experiments. 

\vspace{3mm}
\end{abstract}

\section{Introduction}
High-intensity discharge (HID) lamps are widely used for many lighting purposes. It is estimated that today about $50$ million HID lamps are in operation worldwide. Though HID lamps are increasingly replaced by light-emitting diodes, they will be used and needed in the future (data from \cite{McKinsey.2012}).

\begin{figure}
\centering
\includegraphics[width=0.5\linewidth]{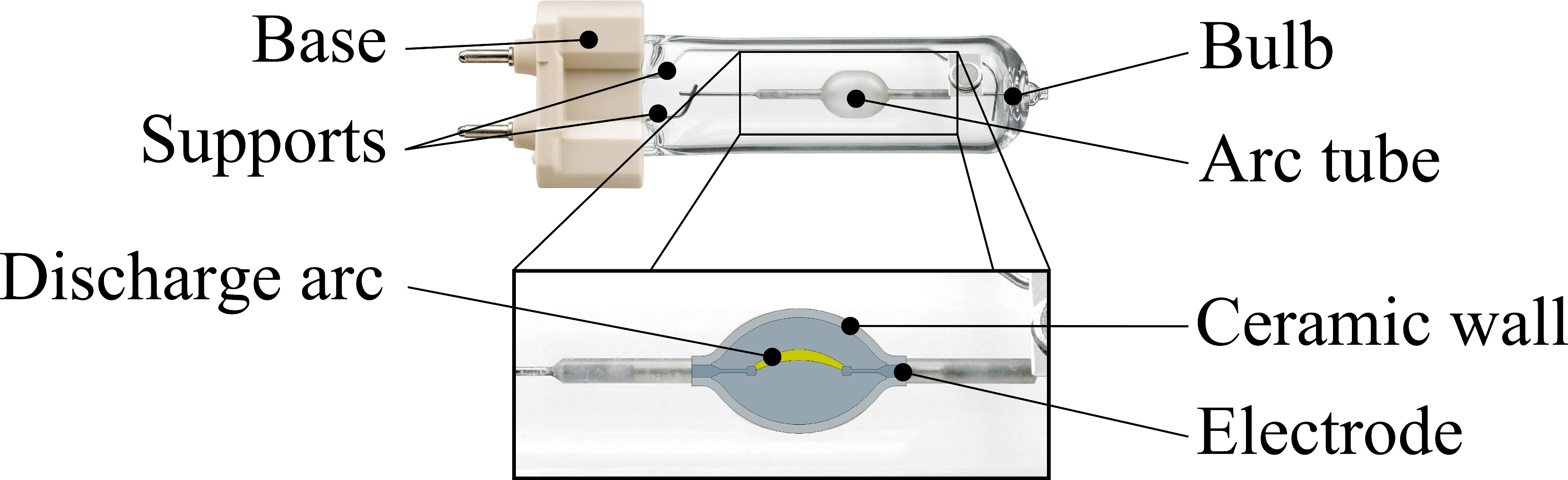}\hspace*{10mm}
\includegraphics[width=0.3\linewidth]{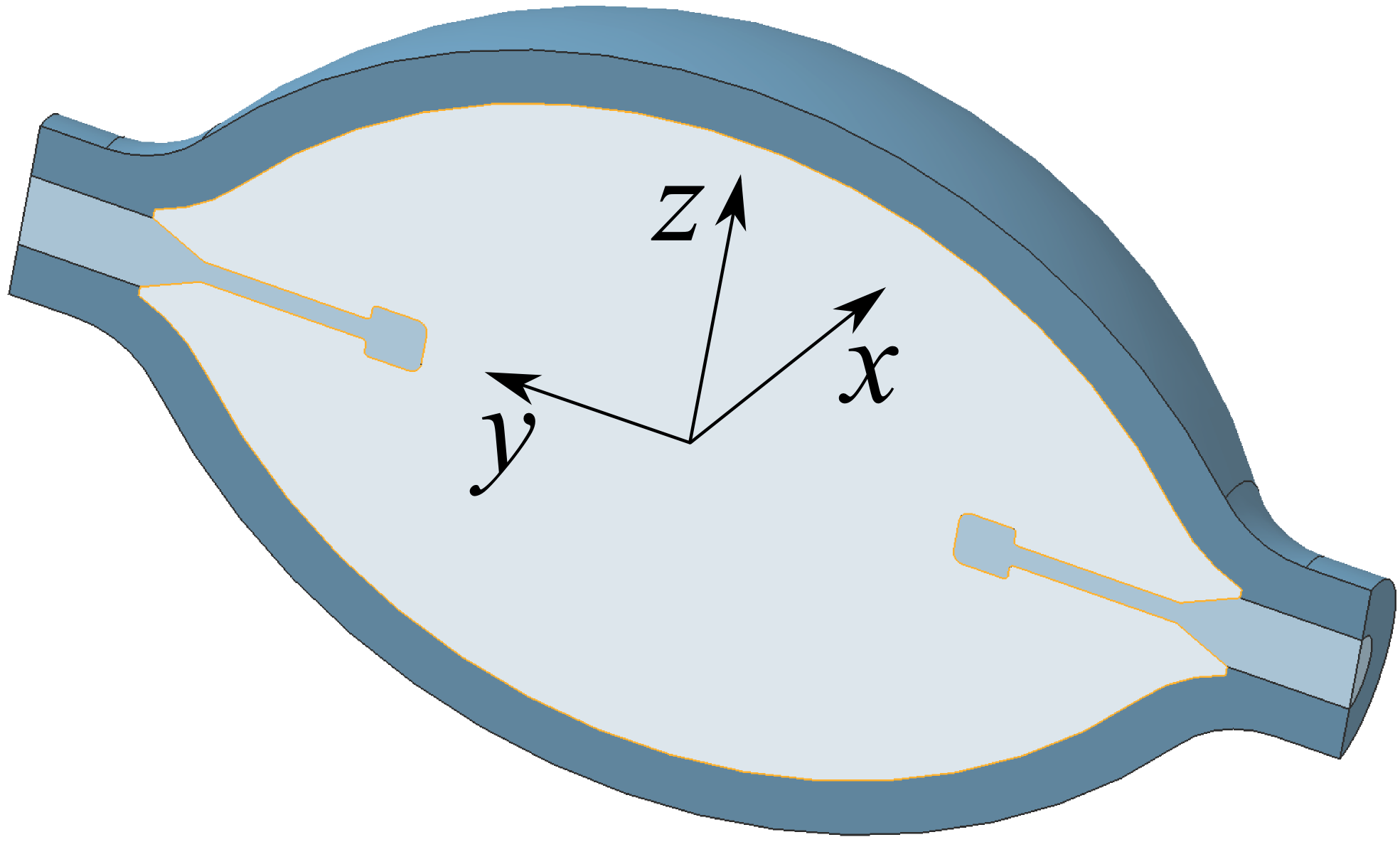}
\caption{Left: Design of an HID lamp. The arc tube is filled with argon, mercury and metal halides. Right: The coordinate system used in this article. The origin is located in the center of the lamp.
\label{fig:Driver2}}
\end{figure}

 The left part of Figure \ref{fig:Driver2} shows the design of the metal halide HID lamp (Philips $\ze{35}{W}$ $930$ Elite), which was investigated. The considered component is the arc tube, respectively its filling. It is disadvantageous concerning cost and efficiency that bulky electronic drivers are needed for lamp operation. In principle, a driver that operates HID lamps at high frequency alternating current (AC) can overcome these drawbacks \cite{Anton.2003}. It has been estimated that an AC frequency of ca. $\ze{300}{kHz}$ would be optimal in this respect \cite{Trestman.2002}. Unfortunately, stable lamp operation for the $\ze{35}{W}$ lamp is not possible at frequencies $\ge\ze{40}{kHz}$. It has been known for a long time that at high frequency operation acoustic resonances are excited inside the arc tube \cite{Witting.1978}. This leads to light flicker, affecting the light quality and the lamp lifetime \cite{Lister.2004}. More recently it has been stated that the acoustic resonances induce a strong flow of the arc tube content via the acoustic streaming (AS) effect, which is the actual source of the lamp instability \cite{Dreeben.2008,Afshar.2008}. 

A crucial step towards a solution of the flicker problem is to understand the  mechanisms, which result in unstable behavior. For this purpose numerical models are very helpful, since these enable to observe the behavior of all physical quantities of interest and the response of the system to parameter changes. For example, it is relatively easy to numerically determine the flow field inside the arc tube and to investigate the influence of design parameters \cite{Schwieger.2013,Okada.1987,Flesch.2003}. Both would be very difficult and elaborate to do experimentally.

Dreeben used an instationary finite element (FE) model to simulate the processes inside the arc tube \cite{Dreeben.2008}. In order to limit memory capacity and computing time, he developed a two dimensional model corresponding to a lamp of infinite length. The results described in the article at hand base on a three dimensional stationary model \cite{0022-3727-48-25-255501,schwieger2015backcoupling}. The idea is to identify the onset of light flicker by the occurrence of instabilities of the stationary solution.

The model has been used to calculate various physical quantities of interest like the temperature field and the acoustic response. Also the voltage drop between the electrodes has been calculated as a function of the driving frequency. The results have been found to be in good accordance with the measured voltage. 

\section{Duffing Oscillator}
\label{sec:DuffingOscillator}

The results described subsequently in this article show a strong resemblance to the nonlinear behavior of the forced Duffing oscillator \cite{duffing1918erzwungene,kalmar2011forced}. Therefore, some characteristics of the Duffing oscillator are presented, which are relevant for the flicker problem in HID lamps. The presentation is mainly inspired by an article of  Kalm{\'a}r-Nagy and Balachandran \cite{kalmar2011forced}.

Considered is an oscillator with a nonlinear spring. The dynamics of the oscillator can be described by the ordinary differential equation
{\nfm{m\ddot{y}+c\dot{y}+k_1y+k_3y^3=F\cos\omega t\label{eq:Duffing0}}
($m$ mass, $y$ displacement, $c$ damping coefficient, $F$ excitation force amplitude, $\omega$ excitation frequency\footnote{Actually, this is the angular frequency.}, $k_1$ linear spring constant, $k_3$ cubic stiffness parameter).
For $k_3=0$ the differential equation describes the behavior of a linear oscillator with an eigenfrequency equal to $\omega_0=\sqrt{\frac{k_1}{m}}$. For $k_3\ne 0$ Equation (\ref{eq:Duffing0}) describes an oscillator, which is characterized by a nonlinear force to displacement relationship. For $k_3>0$ the spring hardens with increasing displacement (\emph{spring hardening}), for $k_3<0$ the spring softens with increasing displacement (\emph{spring softening}).}

{For the following it is convenient to rewrite Equation (\ref{eq:Duffing0}) in non-dimensional form:
\nfm{\ddot{y}+2\zeta\dot{y}+y+\gamma y^3=y_0\cos\Omega \tau.\label{eq:Duffing1}}
This is accomplished by rescaling certain quantities according to
$\zeta:=\frac{c}{2\sqrt{k_1m}}$, $y_0:=\frac{F}{k_1}$ (static amplitude) , $\Omega:=\frac{\omega}{\omega_0}$ and $\tau:=\omega_0t$.}



{Equation (\ref{eq:Duffing1}) does not have a closed form solution for $\gamma\ne 0$ . A qualitative understanding of the amplitude response is obtained by a perturbation analysis: With the aid of a small parameter $\varepsilon\ll1$ the parameters in Equation (\ref{eq:Duffing1}) are expressed as $\zeta=\varepsilon\bar{\zeta}$, $\gamma=\varepsilon\bar{\gamma}$ and {$y_0=\varepsilon\bar{y}_0$}. 
Using the expansion
\nfm{y(\tau)=\varepsilon y_1(\tau)+\varepsilon^2y_2(\tau)+...}
and collecting terms of the same degree in $\varepsilon$ shows that resonances appear at $\Omega\approx~1$ (\emph{primary resonance}), $\Omega\approx 1/3$ (\emph{subharmonic secondary resonance}) and $\Omega\approx 3$ (\emph{superharmonic secondary  resonance}).}

In the following the amplitude response near the primary resonance is investigated. This results in a steady-state solution\footnote{Technical details can be found in \cite{kalmar2011forced}.} with an amplitude response equation of third degree in $a^2$:{
\nfm{{y}^2_0=4a^2\left({\zeta}^2+(\Omega-1-\frac{3}{8}{\gamma}a^2)^2\right).\label{eq:AmpResp1}}}
The resonance frequency is
\nfm{\Omega_{\rm P}=1+\frac{3}{8} \gamma a_{\rm P}^2\label{eq:ResFreq2}}
and the peak amplitude is {
\nfm{a_{\rm P}=\frac{{y}_0}{2{\zeta}}.\label{eq:ResAmpl2}}}

Equation (\ref{eq:ResFreq2}) reveals that for $\gamma<0$ (softening) the resonance frequency is smaller than the eigenfrequency of the linear oscillator, for $\gamma>0$ (hardening) the resonance frequency is larger. The frequency shift increases with the peak amplitude $a_{\rm P}$. For a strong  nonlinearity resonance appears far off the eigenfrequency. Equation (\ref{eq:ResFreq2}) can be recast into {
\nfm{a_{\rm P}(\Omega_{\rm P})=\sqrt{\frac{8(\Omega_{\rm P}-1)}{3\gamma}}.\label{eq:Backbone1}}}
This equation describes the so-called \emph{backbone curve}.

\begin{figure}[tb]
\centering
\includegraphics[width=0.7\linewidth]{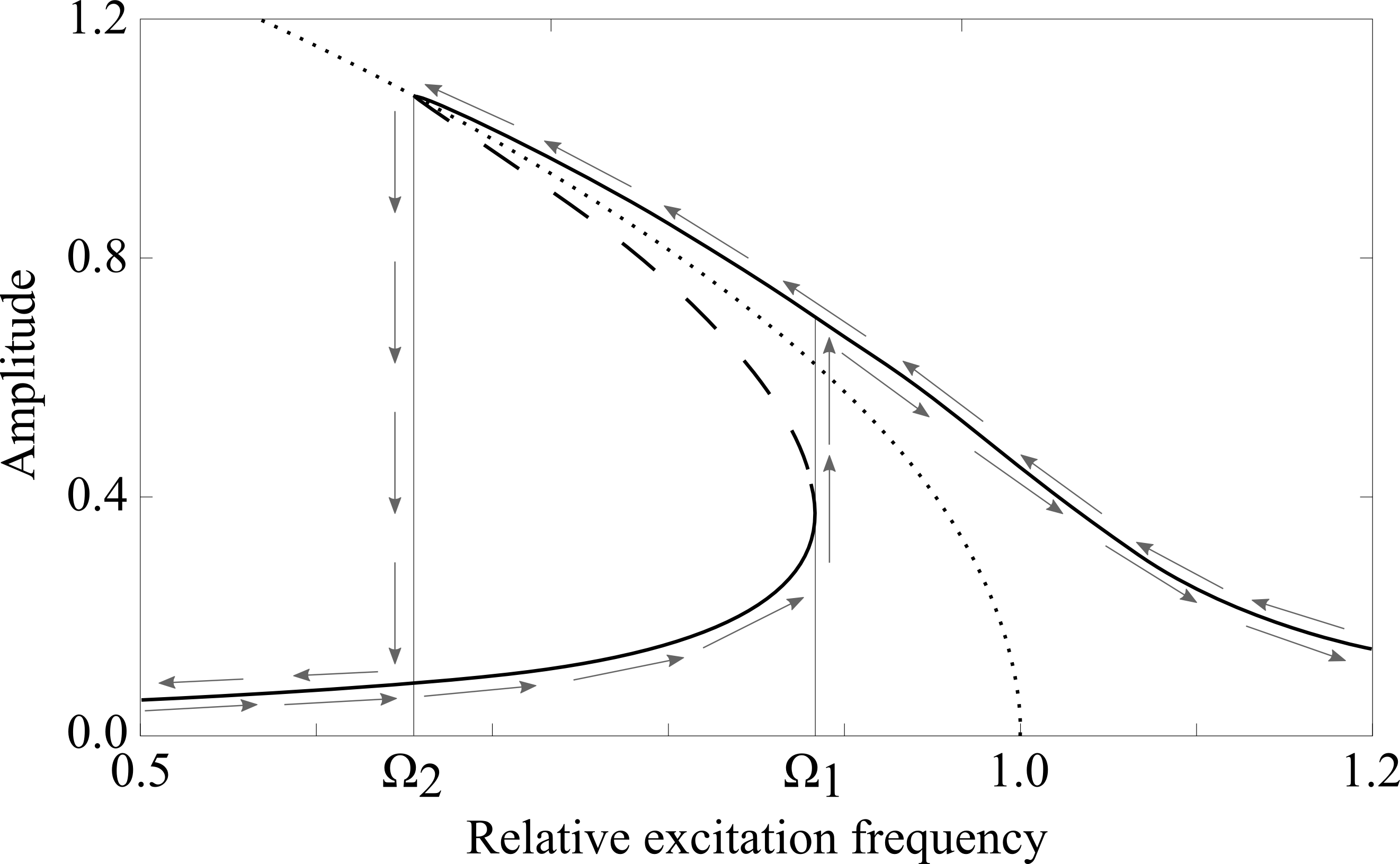}
\caption{{Amplitude response curve of a Duffing oscillator with a softening spring (${y}_0=0.06$, ${\zeta}=0.028$, ${\gamma}=-0.8$).} The backbone curve is depicted by the dotted line. The dashed part of the response curve corresponds to an unstable solution. The arrows indicate the paths of the system during up-/down-ramping in the amplitude-frequency plane.
\label{fig:AmpRespSoft1}}
\end{figure}

Figure \ref{fig:AmpRespSoft1} shows the amplitude response together with the backbone curve of a Duffing oscillator with a softening spring. An interesting phenomenon is observed, when the frequency is increased quasi-statically starting at $\Omega=0$. The amplitude of the oscillation follows the amplitude resonance curve up to the frequency $\Omega_1$, where the curve has a vertical tangent. At $\Omega_1$ the system jumps to the upper branch of the amplitude response curve (\emph{jump phenomenon}). When the frequency is further increased, the amplitude decreases following the upper curve. When the frequency is decreased quasi-statically starting at a high value ($\Omega>1$), a jump from the upper to the lower branch occurs at the frequency $\Omega_2$, where the amplitude reaches its peak value $a_{\rm P}$. The system follows different paths in the $a$-$\Omega$-plane for up- and for down-ramping (\emph{hysteresis}). 

For the intervals $\Omega<\Omega_2$ and $\Omega>\Omega_1$ Equation (\ref{eq:AmpResp1}) has one solution. For the \emph{interval of bistability} $\Omega_2<\Omega<\Omega_1$ three solutions (two stable solutions and one instable solution) exist, and at $\Omega=\Omega_1$ and $\Omega=\Omega_2$ Equation (\ref{eq:AmpResp1}) has two solutions because the instable and one of the stable solutions have merged.

When the static amplitude $y_0$ of the nonlinear oscillator with a softening spring is increased above a critical value, the solution of Equation (\ref{eq:Duffing1}) suffers a \emph{symmetry breaking} or \emph{pitchfork bifurcation}. The pendulum does not oscillate harmonically any more. The movement of the system is then described by one of two possible stable solutions of Equation (\ref{eq:Duffing1}). These solutions are not symmetric with respect to the equilibrium position of the pendulum (symmetry breaking). Further increasing the excitation amplitude results in a \emph{period-doubling cascade} and finally to chaotic solutions \cite{kalmar2011forced}.

\section{Model and Coupling via Recursion}
\label{sec:ModelBackCouplingRecursion}
The FE model used to simulate the processes inside the arc tube has been described in detail in a recent publication \cite{0022-3727-48-25-255501}. Therefore, only an outline of the model is presented. As a new aspect, the influence of AS on the temperature profile inside the arc tube is discussed. The corresponding feedback mechanism was not considered in previous investigations \cite{0022-3727-48-25-255501}, and the focus here is the implementation of a recursion procedure.

\begin{figure}
\centering
{\includegraphics[width=0.7\linewidth,trim=75 85 55 80,clip]{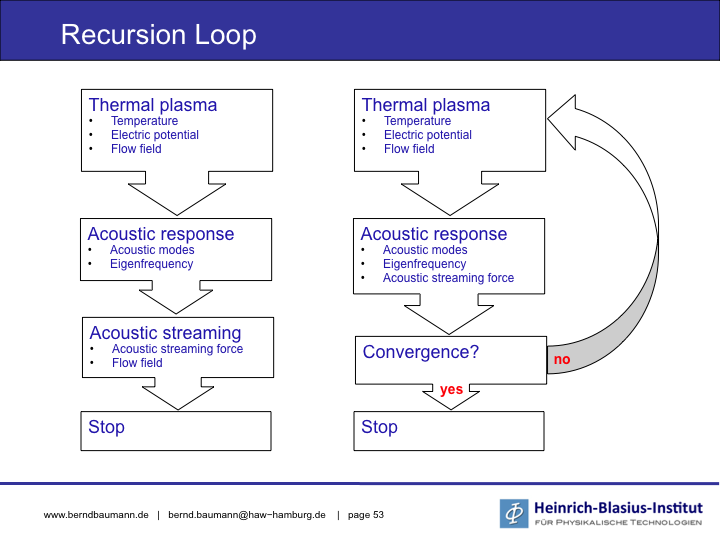}}
\caption{Calculation of the fields without feedback (left) and with feedback via recursion (right). For the calculation of the flow field in the first recursion loop the AS force field is set to zero.
\label{fig:Recursion1}}
\end{figure}

Initially a system of three stationary coupled partial differential equations describing charge, momentum and energy balance subject to appropriate boundary conditions is solved (Figure \ref{fig:Recursion1}, left). The solution of this system  determines the electric potential $\phi(\vec{r})$, the temperature field $T(\vec{r})$ and the buoyancy-driven velocity field $\vec{u}(\vec{r})$ inside the arc tube. On the basis of the temperature profile the acoustic modes are calculated. Using a well-established procedure \cite{Baumann.2009}, we then calculate the acoustic pressure inside the arc tube near and at the specific eigenfrequency. In this step viscous loss and heat conduction loss at the inner wall as well as in the interior of the arc tube are taken into account. It is assumed that the frequency-dependent acoustic pressure can be approximated by a Lorentzian profile \cite{Kreuzer.1977}. This enables to determine the pressure amplitude at a specific AC frequency (excitation frequency). The strong sound waves, which result from the previous step, are responsible for the formation of the AS force field $\vec{F}_{\rm AS}(\vec{r})$. The flow field is now driven not only by the buoyancy force but also by the AS force.

The velocity of the AS field turns out to be considerably higher than the velocity resulting from the buoyancy force. Hence, the AS field will change the temperature field, and the FE model has to be extended in order to take this into account. As this is not directly possible, it is accomplished by a recursion procedure. This recursion comprises the following steps (see Figure \ref{fig:Recursion1}, right):
The fields  $\phi(\vec{r})$,  $T(\vec{r})$, $\vec{u}(\vec{r})$, the acoustic modes and eigenfrequencies as well as $\vec{F}_{\rm AS}(\vec{r})$ are calculated for an excitation frequency far off the considered eigenfrequency as described above. The same fields are calculated once more, but this time  $\vec{F}_{\rm AS}(\vec{r})$ from the previous iteration step is added to the buoyancy force. Accordingly, the resulting AS force field will change. All fields are repeatedly calculated until convergence. As a convergence criterion, the eigenfrequencies of the current and the previous recursion step are compared. If these differ by less than $\ze{10}{Hz}$, convergence is assumed. Then the excitation frequency is changed to a value closer to the resonance frequency, and the complete procedure is repeated using the converged solution as initial condition.

\section{Results}
\label{sec:Results}
\subsection{Acoustic Modes}
\label{sec:AcousticModes}
\begin{figure}
\centering
\includegraphics[width=0.32\linewidth]{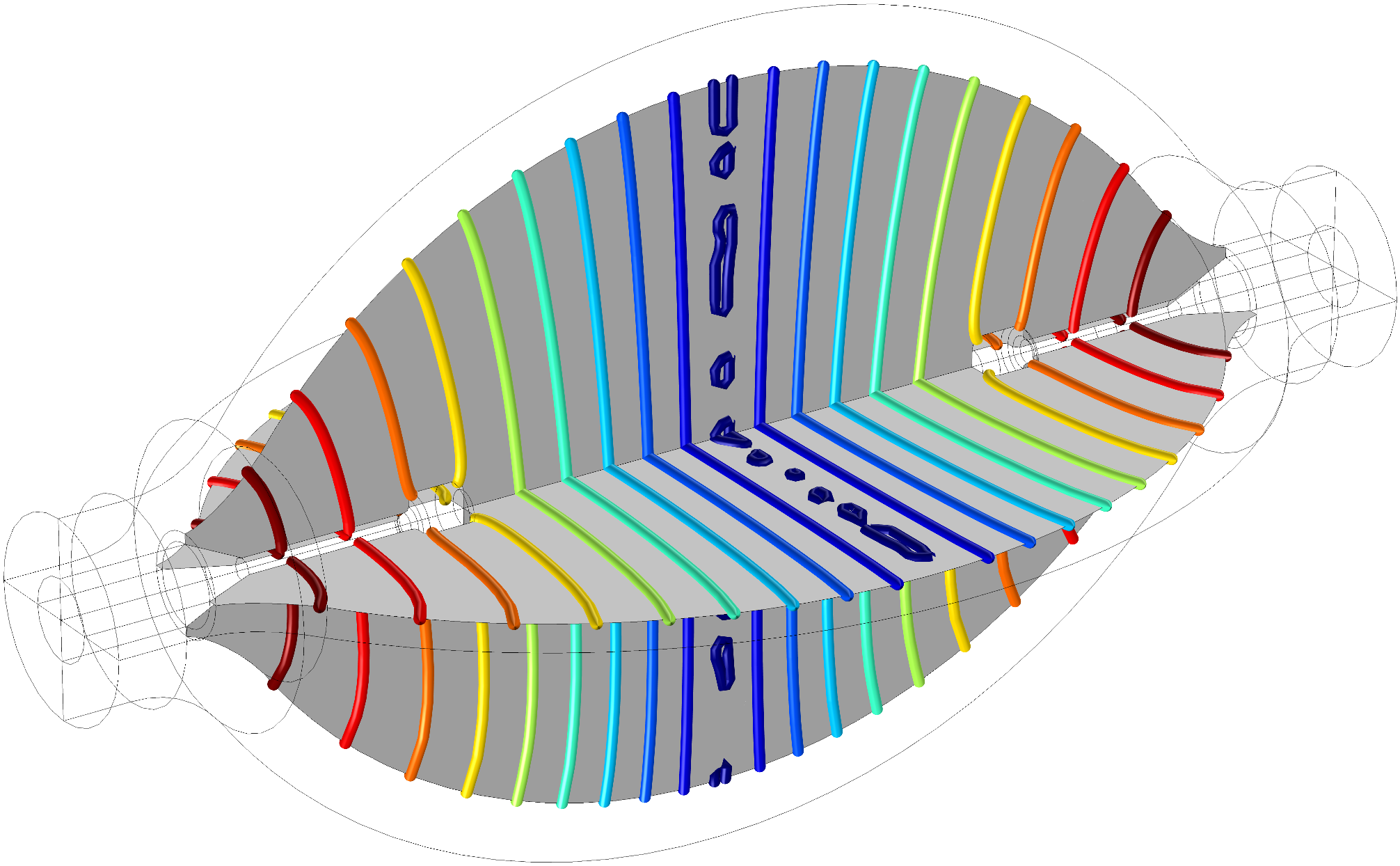}
\includegraphics[width=0.32\linewidth]{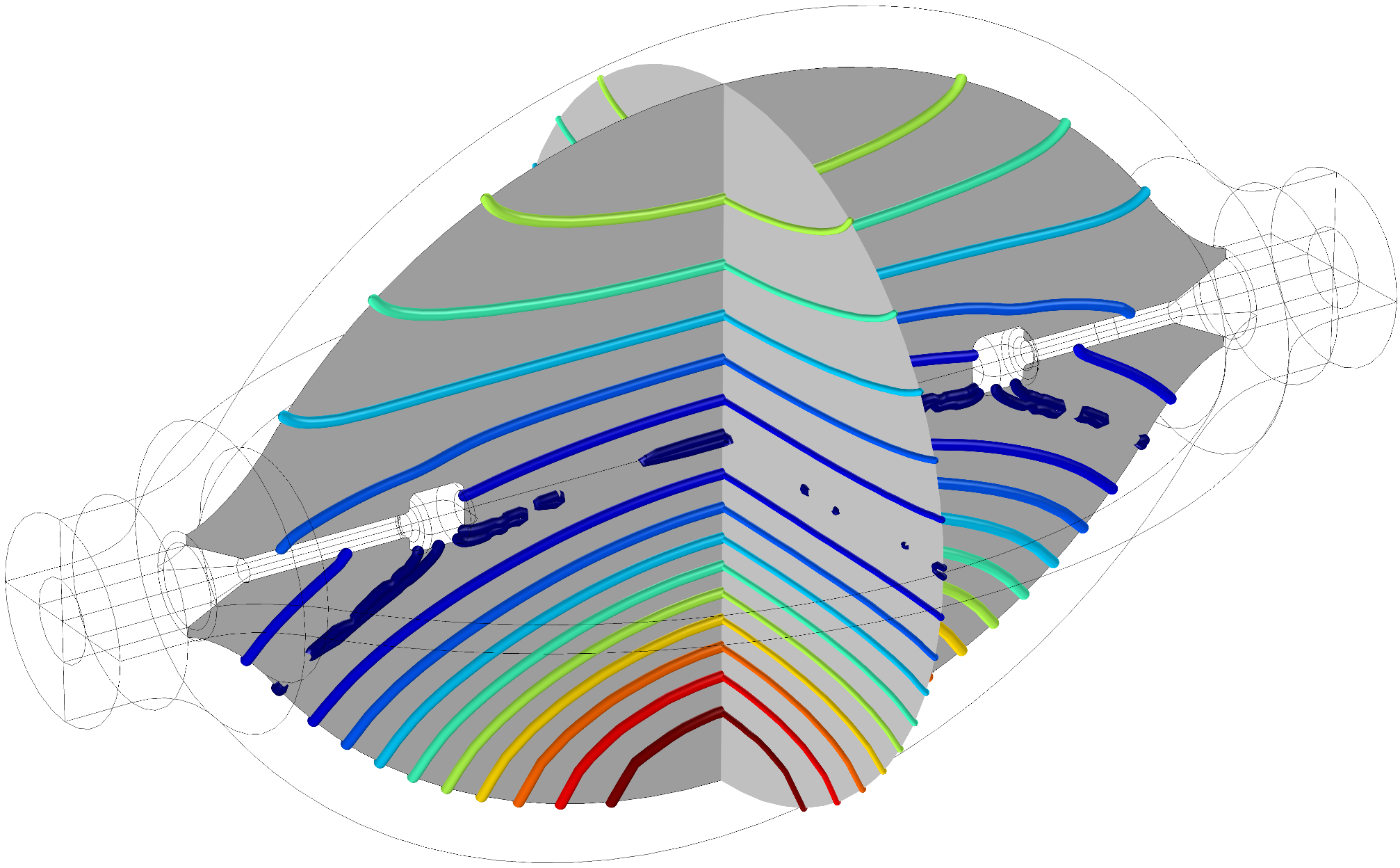}
\includegraphics[width=0.32\linewidth]{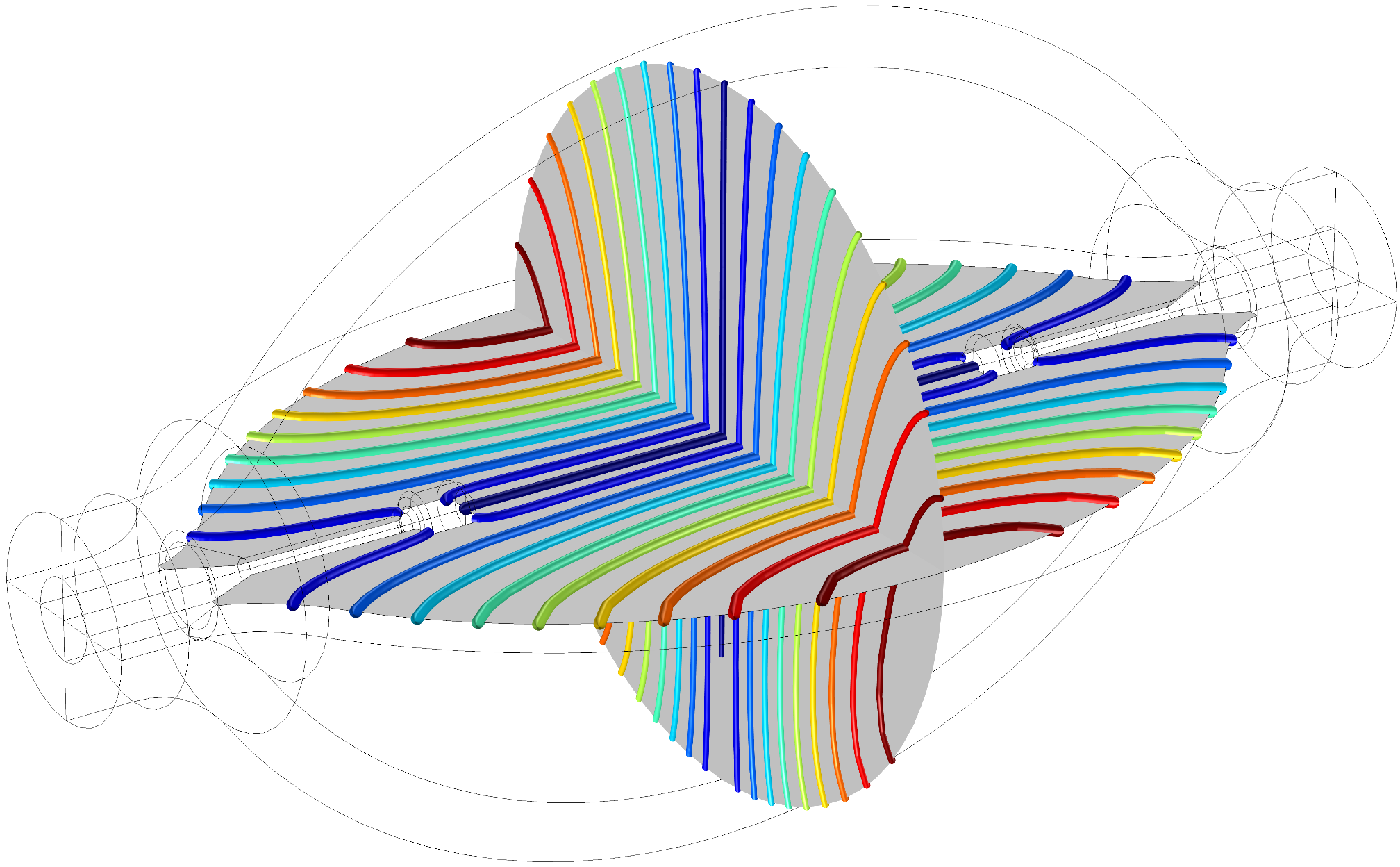}
\caption{The three lowest acoustic modes resulting from the FE model. Depicted is $|p|$, where $p$ is the acoustic pressure. Left: $f=\ze{33.1}{kHz}$, middle: $f=\ze{47.8}{kHz}$, right: $f=\ze{48.1}{kHz}$. Blue indicates $|p|\approx 0$ and red the maximum of $|p|$.
\label{fig:AcEigenmodes1}}
\end{figure}

For the lamp considered in this investigation the three acoustic modes with the lowest eigenfrequencies are depicted in Figure \ref{fig:AcEigenmodes1}. No signature of the first mode at $f=\ze{33.1}{kHz}$ is found in the experiments described in Section \ref{sec:VoltageDrop}. 
This can be attributed to the fact that the excitation amplitude of the $j$-th acoustic mode is proportional to an integral over the interior of the arc tube \cite{Kreuzer.1977}
\nfm{
\int_{}{}{p_j^\ast {\cal H}}{{\;\rm d}V.}\label{eq:excAmplitude}}
For the first mode the overlap integral of the power density of heat generation ${\cal H}(\vec{r})$ and the conjugate complex of the mode's acoustic pressure distribution $p_j^\ast (\vec{r})$ is very small. The power density is large inside the plasma arc, i.e.\ between the electrodes, where the mode with the lowest frequency has a pressure node, as it is clearly visible in the left part of Figure~\ref{fig:AcEigenmodes1}. 

The second and the third mode correspond to almost equal eigenfrequencies. Actually, in the absence of gravity these two modes are degenerate. The degeneracy is lifted due to buoyancy, which is responsible for an asymmetry with respect to rotations around the $y$-axis. The hot plasma arc bows upward leading to a non-vanishing overlap with the pressure antinode of the second mode (Figure \ref{fig:AcEigenmodes1}, center). As this mode can be experimentally excited, our investigations concern this mode. The third mode is not observed in experiments for the same reason as the first mode: The overlap of the plasma arc and the antinode is small.

\subsection{Acoustic Response}
\label{sec:AcousticResponse}
\begin{figure}
\centering
\includegraphics[width=0.5\linewidth]{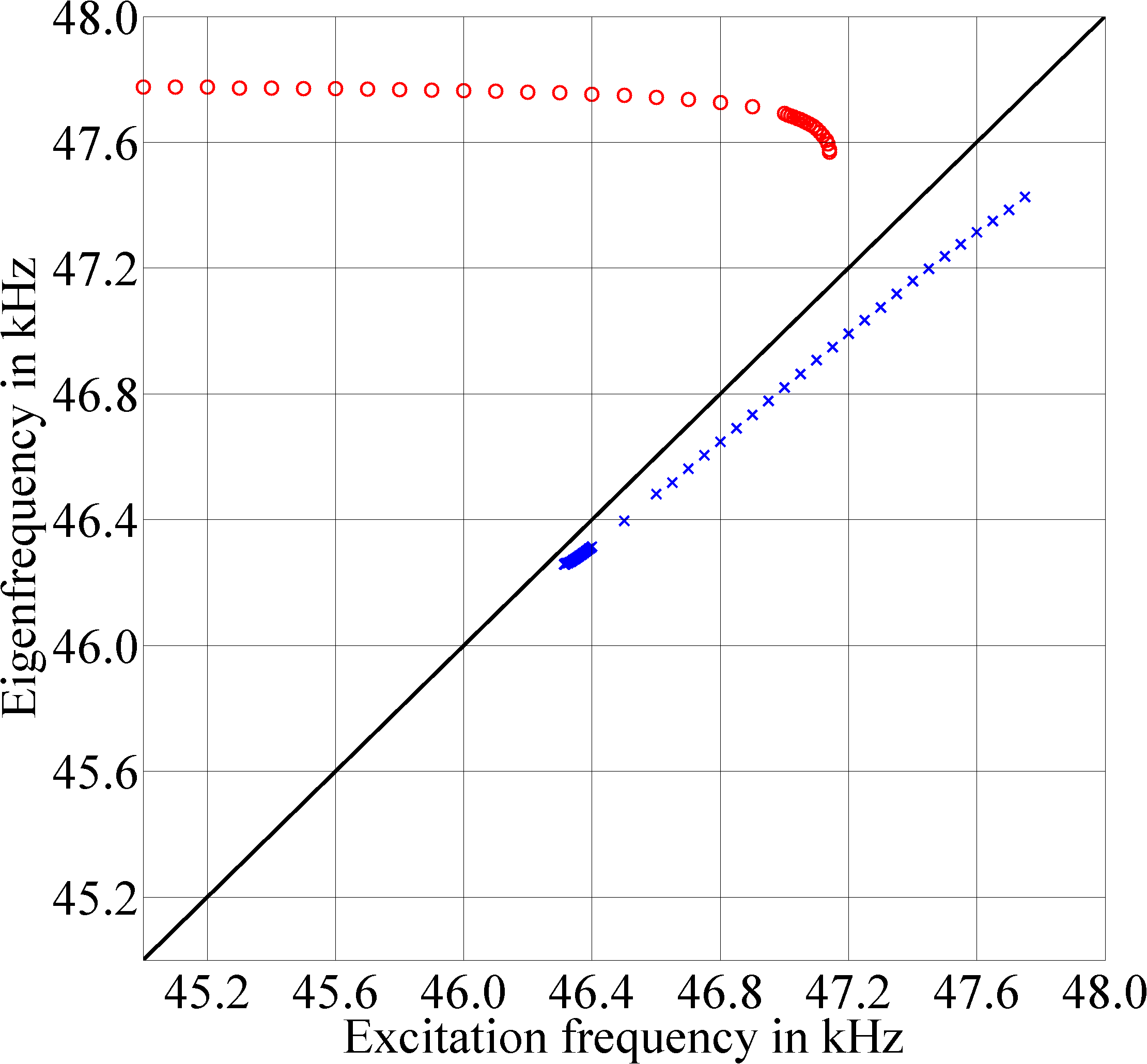}
\caption{The acoustic eigenfrequencies vs. the excitation frequency. Blue crosses correspond to decreasing excitation frequencies, red circles to increasing excitation frequencies. At the black line excitation frequencies and eigenfrequencies are equal (resonance). 
\label{fig:EFvsXF}}
\end{figure}

In this section results of the FE model for the acoustic response are presented. As described in Section \ref{sec:ModelBackCouplingRecursion}, a computing scheme has been adopted to approach the eigenfrequency of the second mode. In Figure \ref{fig:EFvsXF} this process is summarized. When trying to approach the eigenfrequency by decreasing the excitation frequency (blue crosses), the system behaves as if the eigenfrequency is repelled from the excitation frequency. The conditions inside the arc tube change in a way that  the eigenfrequencies are shifted to smaller values. At around $\ze{46.4}{kHz}$, where excitation frequency and eigenfrequency lie very close together, this process decelerates. At the same time the convergence of the recursion deteriorates, i.e.\ significantly more recursion loops are required for convergence. For this reason the step size of the frequency change was reduced. At an excitation frequency of $\ze{46.3}{kHz}$ no convergence was achieved. The iteration steps of the simulation show acoustic pressure jumps between two values.


When approaching the eigenfrequency by increasing the excitation frequency (red circles), a quite different behavior is observed. The eigenfrequency hardly changes when increasing the excitation frequency. At about $\ze{47.0}{kHz}$ convergence slows down, and the eigenfrequency gradually decreases. At even higher excitation frequencies the recursion does not converge any more. Once again, pressure jumps between two values are observed.

\begin{figure}
\centering
\includegraphics[width=0.9\linewidth]{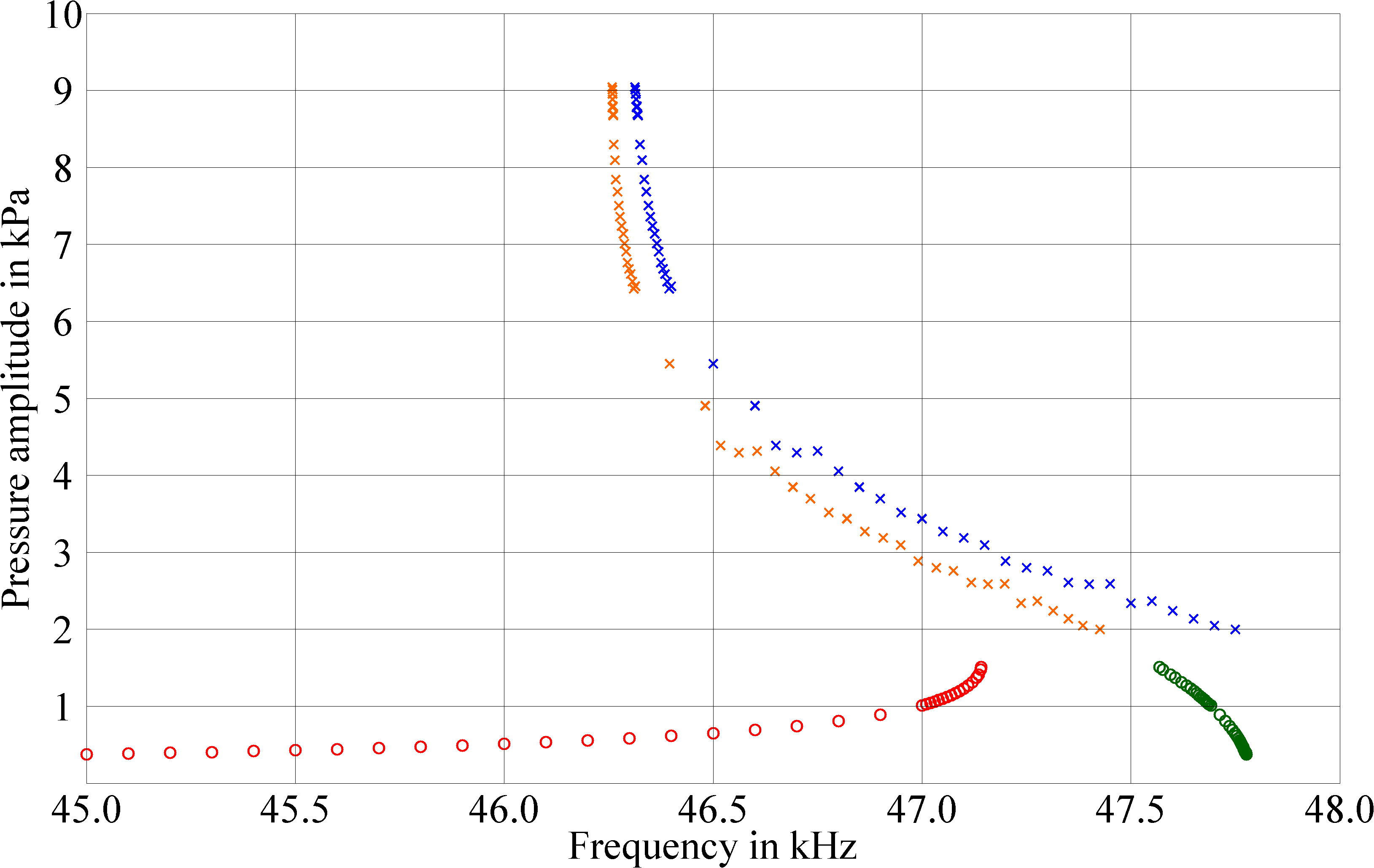}
\caption{Acoustic response vs. excitation frequency for up-ramping (red circles) and down-ramping (blue crosses). In addition the backbone curve (acoustic response vs. eigenfrequency) on the basis of the same pressure data is shown. The green circles indicate up-ramping, the orange crosses indicate down-ramping.
\label{fig:PressVsXFEF}}
\end{figure}

In Figure \ref{fig:PressVsXFEF} the acoustic response, i.e.\ the acoustic pressure amplitude at an antinode as function of the excitation frequency, is depicted. In addition to the up- and down-ramping data as a function of the excitation frequency, the same pressure data are depicted as a function of the eigenfrequency. This was done by utilizing the data underlying Figure \ref{fig:EFvsXF}, which connect excitation frequency and eigenfrequency. Under the reasonable assumption that the resonance frequency and the eigenfrequency are very close to each other, the curve can be interpreted as the backbone curve (see Section \ref{sec:DuffingOscillator}). The emerging picture resembles the situation of the Duffing oscillator depicted in Figure \ref{fig:AmpRespSoft1}. Naturally, the branch representing the unstable solution does not appear in the simulation result. The jumps of the acoustic pressure mentioned above can now be related to the jump phenomenon described in Section \ref{sec:DuffingOscillator}. The jumps occur between the upper and the lower branch of the acoustic response curves. The frequencies, where the simulations do not converge any more, are related to $\Omega_1$ ($f_1=\Omega_1/2\pi=\ze{47.141}{kHz}$) and $\Omega_2$ ($f_2=\Omega_2/2\pi=\ze{46.313}{kHz}$) - the boundaries of the interval of bistability.

For a softening spring the force for large displacements  becomes repulsive, and, therefore, is not appropriate to describe the behavior of a real spring. The acoustic response curve of Figure \ref{fig:PressVsXFEF} avoids this unphysical behavior by steepening. The vertical tangent at high acoustic pressure values indicates that the softening effect is saturated. The modulus of compression (inverse of the compressibility) is the acoustical analogue of the spring constant of a mechanical oscillator. The modulus of compression increases to a new constant value when approaching the eigenfrequency.

\subsection{No Symmetry Breaking}
\label{sec:NoSymmetryBreaking}
\begin{figure}
\centering
\includegraphics[width=0.9\linewidth]{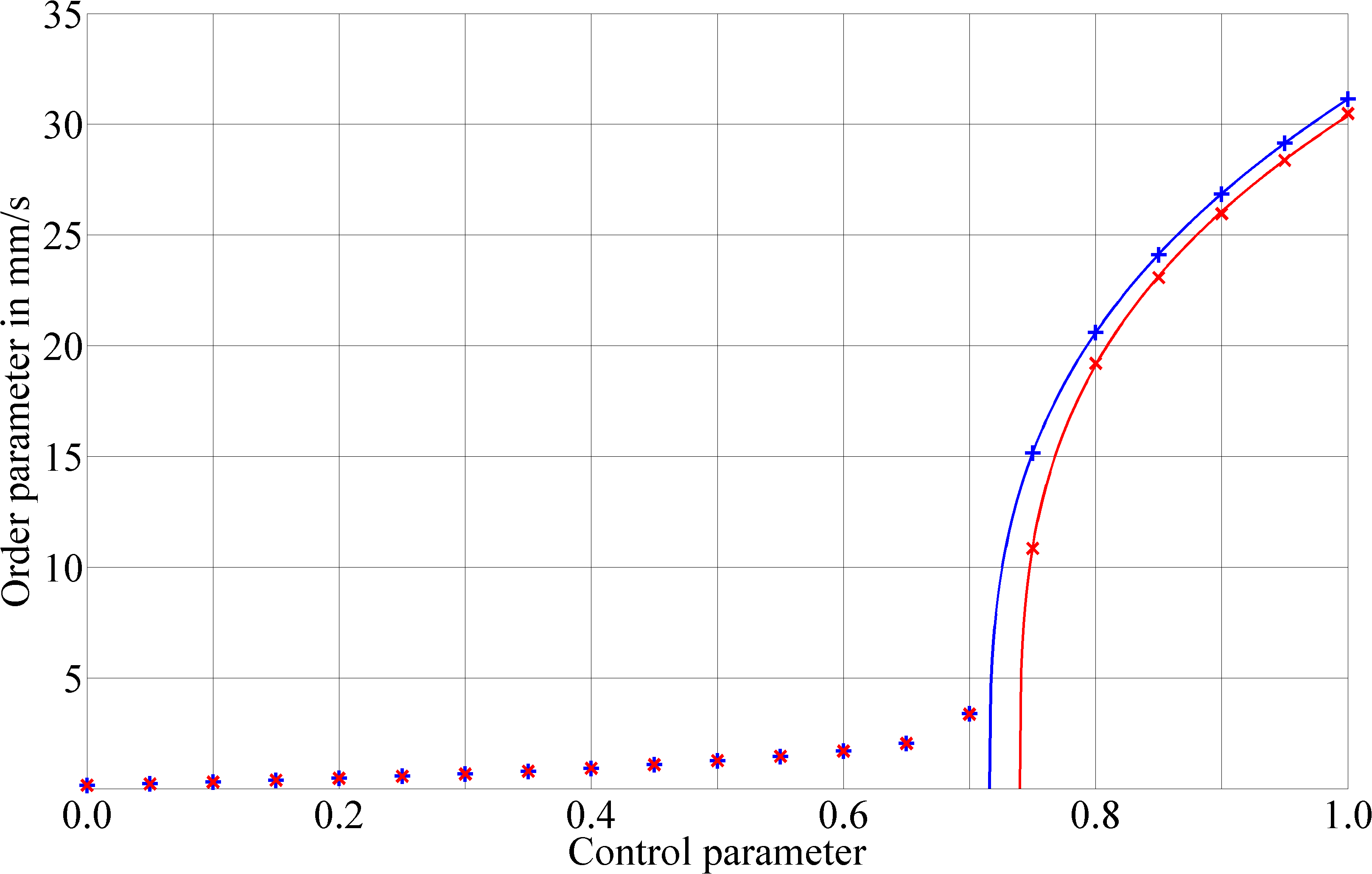}
\caption{Order parameter $\Phi(S)$ for mirror symmetry as a function of the control parameter $S$. The curves result from a simulation without back coupling. Both curves correspond to horizontal lamp operation. The two curves have been obtained from approaching the horizontal lamp state from positive/negative tilt angle.
\label{fig:OrderParameter_ZeroDegree_3}}
\end{figure}

In Section \ref{sec:DuffingOscillator} it was mentioned that a nonlinear oscillator with a softening spring shows a symmetry breaking bifurcation when the excitation amplitude $F$ exceeds a critical value. We previously reported the detection of a symmetry breaking transition in our simulation results \cite{0022-3727-48-25-255501}. For this we introduced a control parameter $S$, which allowed us to continuously interpolate between an AS force equal to zero ($S=0$) and a fully activated AS force ($S=1$). Furthermore, we defined an order parameter $\Phi(S)$, which measured the deviation from mirror symmetry of the flow field inside the arc tube with respect to the $x$-$z$-plane. Figure \ref{fig:OrderParameter_ZeroDegree_3} highlights that for a horizontally operated lamp symmetry breaking sets in for $S$ slightly above $0.7$ (critical point). In order to avoid convergence problems related to the symmetric solution, which becomes unstable at the critical point, a limiting procedure has been adopted, i.e.\ the simulations were started for a tilted lamp and the tilting has been removed gradually.

The FE model described in \cite{0022-3727-48-25-255501} does not consider the feedback of AS on the temperature field. In the present article the back coupling effect is taken into account for a tilt angle of $5\grad$. Thus, the tedious limiting procedure is avoided, and the essential features can still be captured. In addition, a slight tilting has the benefit of being a more realistic scenario for real lamp operation.

The AS force in the lower centrical part of the arc tube points into the opposite direction as the buoyancy force. In \cite{0022-3727-48-25-255501} an order parameter
\nfm{\Psi:=\min\limits_{z\in\mathcal{D}} u_z(0,0,z)}
was introduced. $\mathcal{D}$ is the part of the $z$-axis inside the arc tube and $u_z(x,y,z)$ the $z$-component of the flow velocity. Hence, $\Psi<0$ indicates that somewhere along the $z$-axis a downward directed AS force exists that is larger than the upward-directed buoyancy force at the same point.

\begin{figure}
\centering
\includegraphics[width=0.9\linewidth]{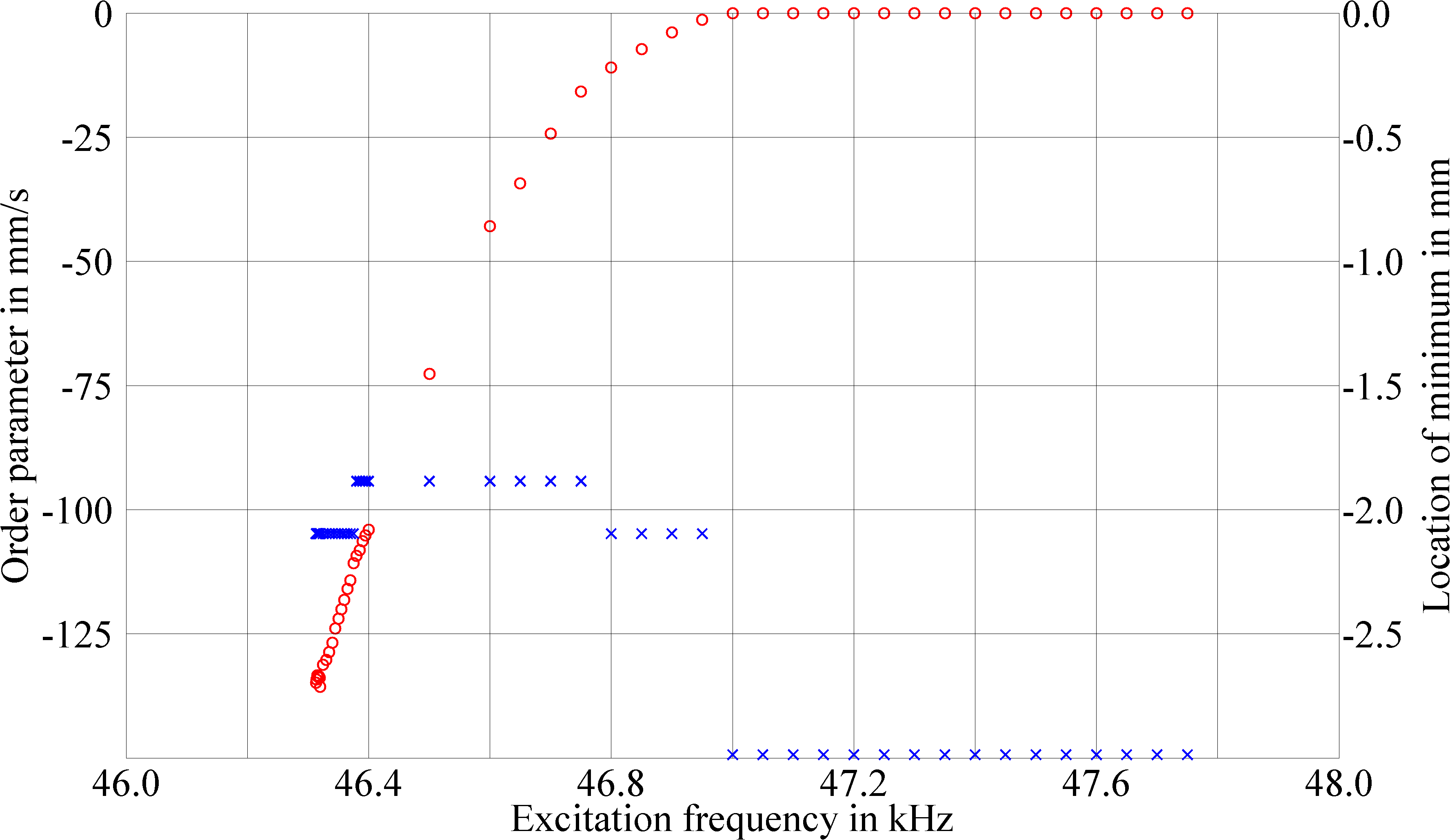}
\caption{Order parameter $\Psi$ vs. excitation frequency $f_{\rm e}$ for descending frequency (red circles). The corresponding axis is on the left. The blue crosses show at which point on the $z$-axis the minimal value of $u_z$ was found (axis on the right). In a buoyancy dominated flow $\Psi$ is equal to zero corresponding to the obvious fact that the fluid cannot penetrate the wall and, therefore, $u_z(0,0,\ze{-3}{mm})=0$. 
\label{fig:PsiVsFreqAsc}}
\end{figure}

The results of the previous investigation show that symmetry breaking is induced by the downward directed flow with a minimal strength of  $\Psi\approx\ze{170}{mm/s}$ \cite{0022-3727-48-25-255501}. In the back coupled model, the flow field does not show symmetry breaking. $\Phi$ has small contributions from the slight tilting and from numerical noise only. Instead of the artificial control parameter $S$, we use the difference between the excitation frequency and the eigenfrequency as a natural control parameter for the strength of the AS force. For up-ramping $\Psi$ is identical to zero for all frequencies and, consequently, the impact of the AS is insignificant. Figure \ref{fig:PsiVsFreqAsc} shows $\Psi(f_{\rm e})$ for down-ramping. AS becomes important for $f_{\rm e}\le\ze{47}{kHz}$. The maximal flow velocity in downward direction in the back coupled model is about $\ze{140}{mm/s}$ only. We conclude that due to back coupling the AS force is reduced and not strong enough to induce symmetry breaking. The system avoids symmetry breaking by the jump phenomenon (Section \ref{sec:DuffingOscillator}).

\subsection{Voltage Drop: Simulation vs. Measurement}
\label{sec:VoltageDrop}
In this section we confront the FE model with experiments. In contrast to the acoustic pressure inside the arc tube, the voltage drop between the electrodes is readily available in the simulation results and can easily be measured experimentally as well. 

\begin{figure}
\centering
\includegraphics[width=0.9\linewidth]{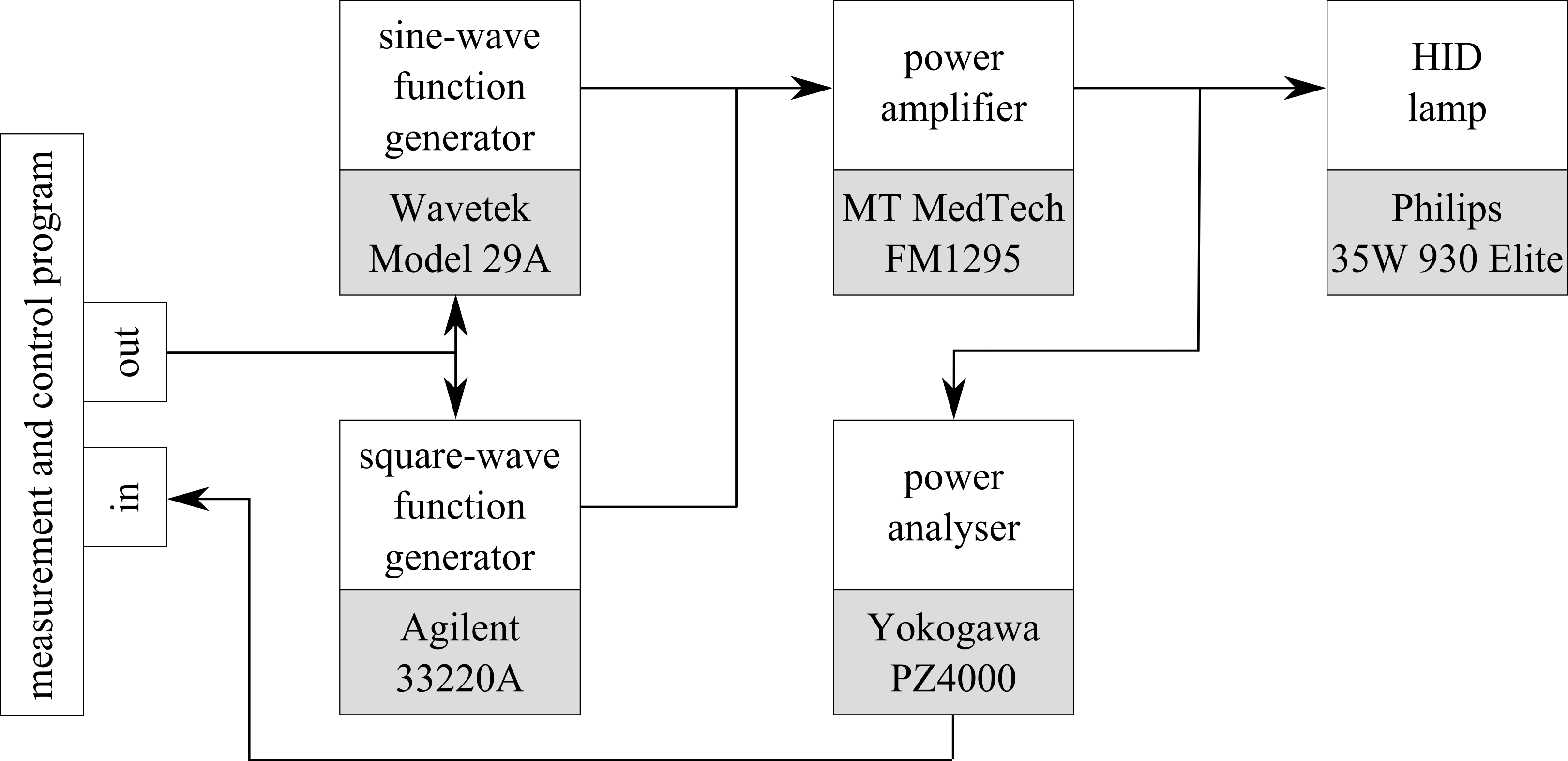}
\caption{Schematic description of the experimental set-up used for the measurement of the voltage drop between the electrodes of the lamp.
\label{fig:SchematicExperimentalSetup}}
\end{figure}

In the experimental set-up (Figure \ref{fig:SchematicExperimentalSetup}) the lamp is supplied with a $f_{\rm c}=\ze{400}{Hz}$ rectangular voltage input, which guarantees stable operation. In order to induce light flicker, a high frequency sinusoidal voltage is superimposed to the square wave input. The input signal can be described by
\fm{U(t) = \begin{cases} A\left[+1+\alpha\sin \left( \omega_{\rm e} t \right) \right] \hspace{1cm} 2n<\omega_{\rm c}t\leq2n+1 \\ A\left[-1+\alpha\sin \left( \omega_{\rm e} t \right) \right] \hspace{1cm} 2n+1<\omega_{\rm c}t\leq2n+2 \end{cases} n\in\mathbb{N}_0,}
where $\omega_{\rm e}=2\pi f_{\rm e}$ and $f_{\rm e}$ is the excitation frequency. The amplitude $A$ is adjusted such that the lamp is operated at the nominal power of $\ze{35}{W}$. The modulation depth $\alpha$ is a measure of the relative contribution of the high frequency voltage to the rectangular voltage input.

The measurements are started by operating the lamp at $\alpha=0$ for at least $20$~minutes. In case of up-ramping the excitation frequency $f_{\rm e}$ is set thereafter to $\ze{32}{kHz}$ and the modulation depth to some fixed value. Then the voltage is measured $40$ times within a period of $\ze{20}{s}$. The mean of these $40$ voltage measurements is called basic voltage, since it corresponds to a frequency far off the active acoustic mode. Next, the excitation frequency is increased by $\ze{50}{Hz}$ and again $40$ voltage measurements within $\ze{20}{s}$ are carried out. This routine is repeated until the excitation frequency has reached $\ze{52}{kHz}$
 or one of two termination criteria is met: If one of the voltage measurements results in $\ze{8}{V}$ or more above the basic voltage, the measurement is stopped in order to avoid lamp failure. For the same reason the measurements are terminated if voltage fluctuations of $\ze{1.5}{V}$ or more are encountered. In the case of down-ramping the same procedure is applied starting at $\ze{52}{kHz}$.
 
\begin{figure}
\centering
\includegraphics[width=0.9\linewidth]{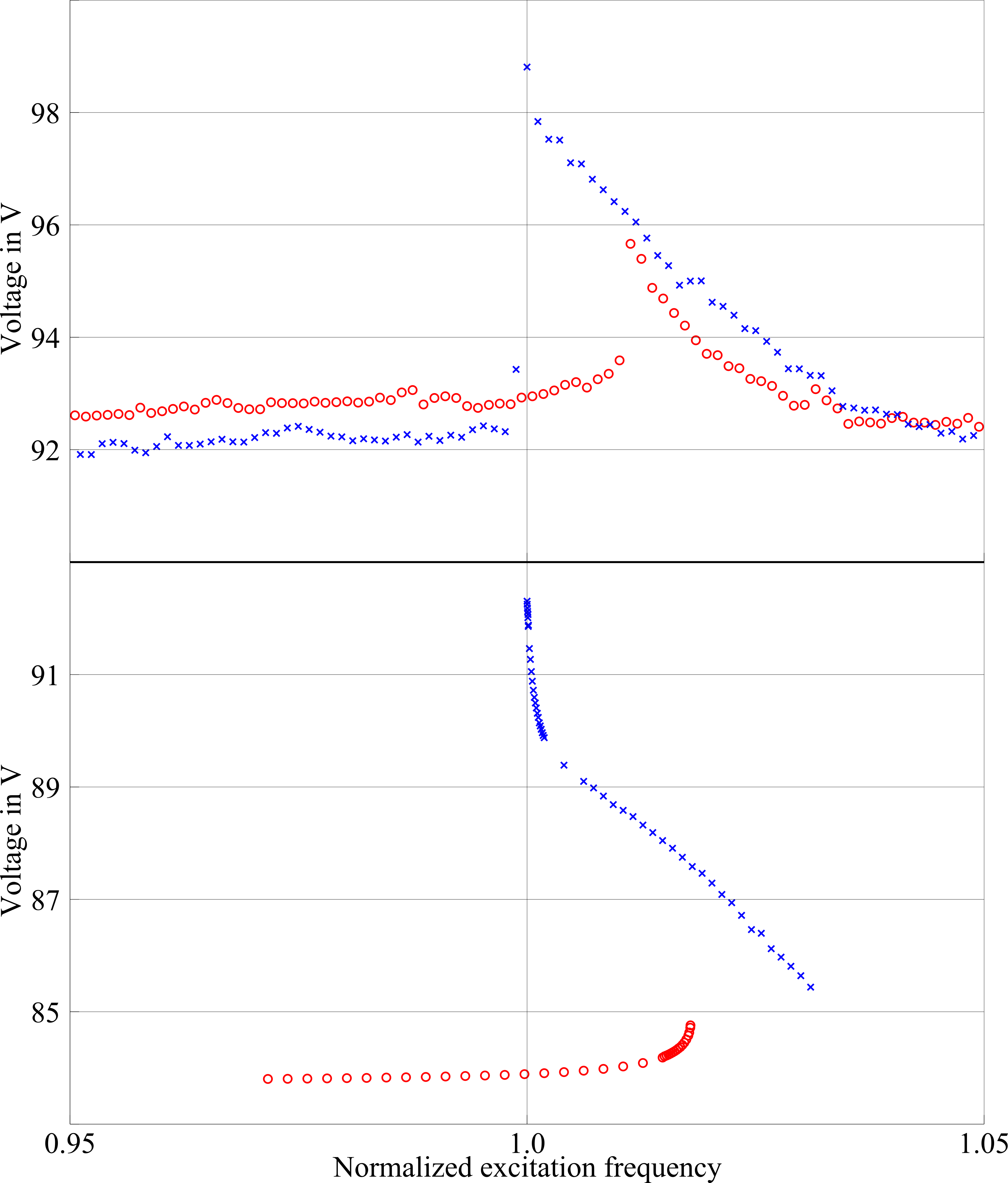}
\caption{Voltage drop between the electrodes. Upper part: Measurement results at a modulation depth of $2\%$. Lower part: Results from the simulation. The normalized excitation frequency is defined by $(f_{\rm e}-\hat{f})/\hat{f}$ (see text). Red circles denote up-ramping, blue crosses down-ramping.
\label{fig:VoltvsXFSimExp}}
\end{figure}

The upper part of Figure \ref{fig:VoltvsXFSimExp} shows the measured voltage for up- and down-ramping at a modulation depth of $2\%$. Each of the circles/crosses represents the mean value of the voltage taken over $6$ individual lamps. The voltage varies considerably from lamp to lamp (standard deviation: ca. $\ze{2.6}{V}$). The deviations of the excitation frequency $\hat{f}$, at which the maximal voltage occurs, varies from lamp to lamp (mean value: $\ze{41.4}{kHz}$, standard deviation: ca. $\ze{0.4}{kHz}$). For this reason the voltage has been plotted against the normalized frequency $(f_{\rm e}-\hat{f})/\hat{f}$. Obviously, the voltage behaves very similar to the acoustic response of Figure \ref{fig:PressVsXFEF}. In particular, the jump phenomenon and the hysteresis are noticeable. The lower part of Figure \ref{fig:VoltvsXFSimExp} shows the simulation results of the voltage. The region of bistability is somewhat wider compared to the measurement results. The voltage as well as the peak frequencies differ by approximately $10\%$, but the general behavior is identical. In view of the approximations made during the modelling and the rather large uncertainties in the material properties the accordance of the results is very satisfactory.


The results of the simulation correspond to a modulation depth of $100\%$. It follows that the FE model underestimates the impact of AS on the lamp behavior. It is not completely clear what the reason for this discrepancy is. An explanation might be that the electrodes in the model are simplified as cylinders to limit the number of degrees of freedom of the FE mesh. Additionally, the temperature-dependent material parameters are a source of error because these are difficult to determine experimentally. Both effects result in a too low temperature field in the simulation compared to real lamp operation. Consequently, a too small AS force is computed. Another possible reason for the difference is that the simulation does not include changes of the plasma composition. When the minimal plasma temperature (cold spot temperature) increases (see Section \ref{sec:TemperatureProfile}), additional salts are evaporated leading to altered material functions that are not considered by the simulation. 

\subsection{Velocity and Temperature Fields}
\label{sec:TemperatureProfile}

In this section we investigate the behavior of certain quantities at different excitation frequencies in order to obtain a better understanding of the processes inside the arc tube. In particular, we take a closer look at the flow and the temperature field. Both quantities are available in the simulation but are not easily accessible in experiments.

\begin{figure}
\centering
\includegraphics[width=0.7\linewidth]{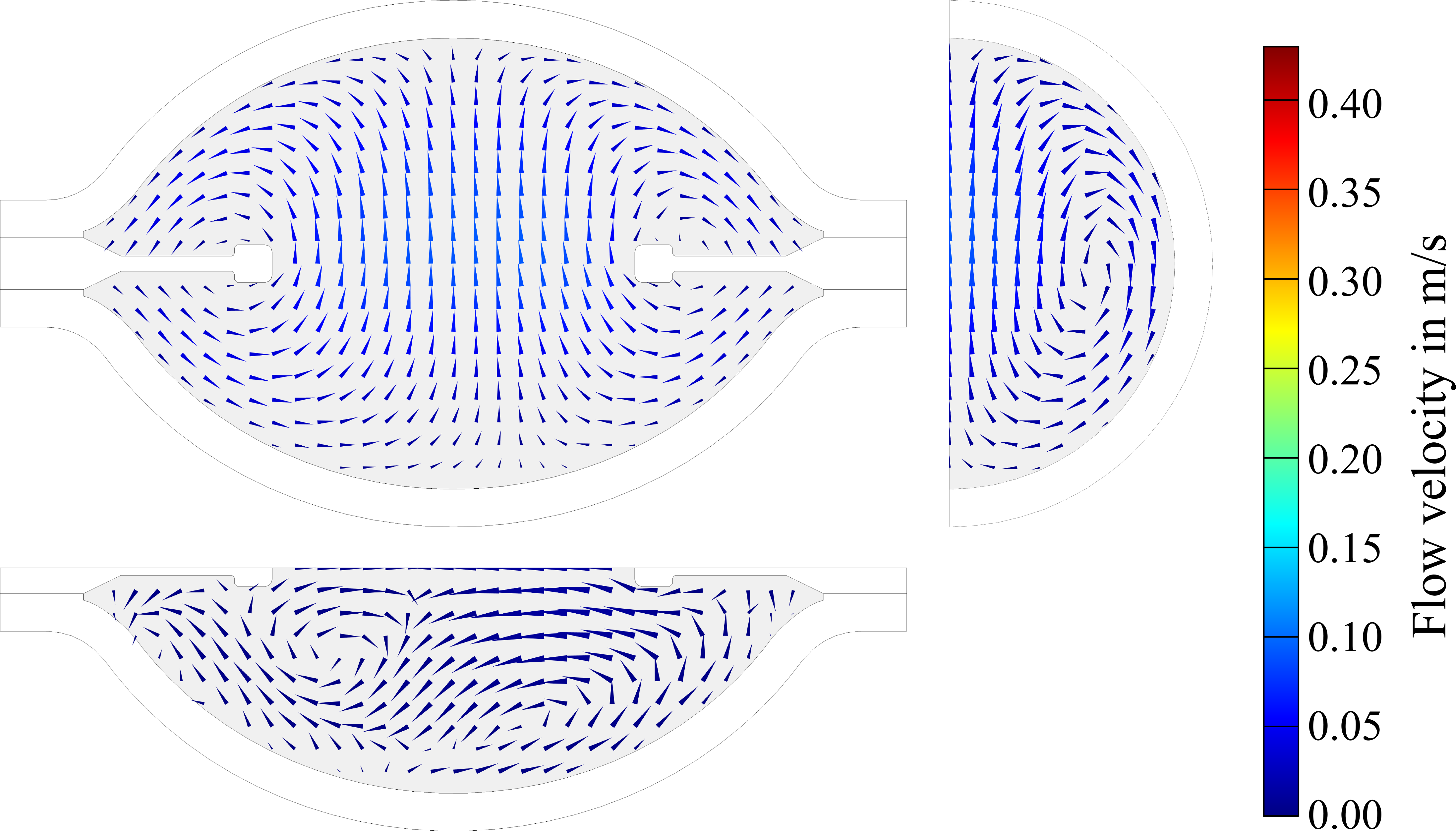}
\vspace*{5mm}
\includegraphics[width=0.7\linewidth]{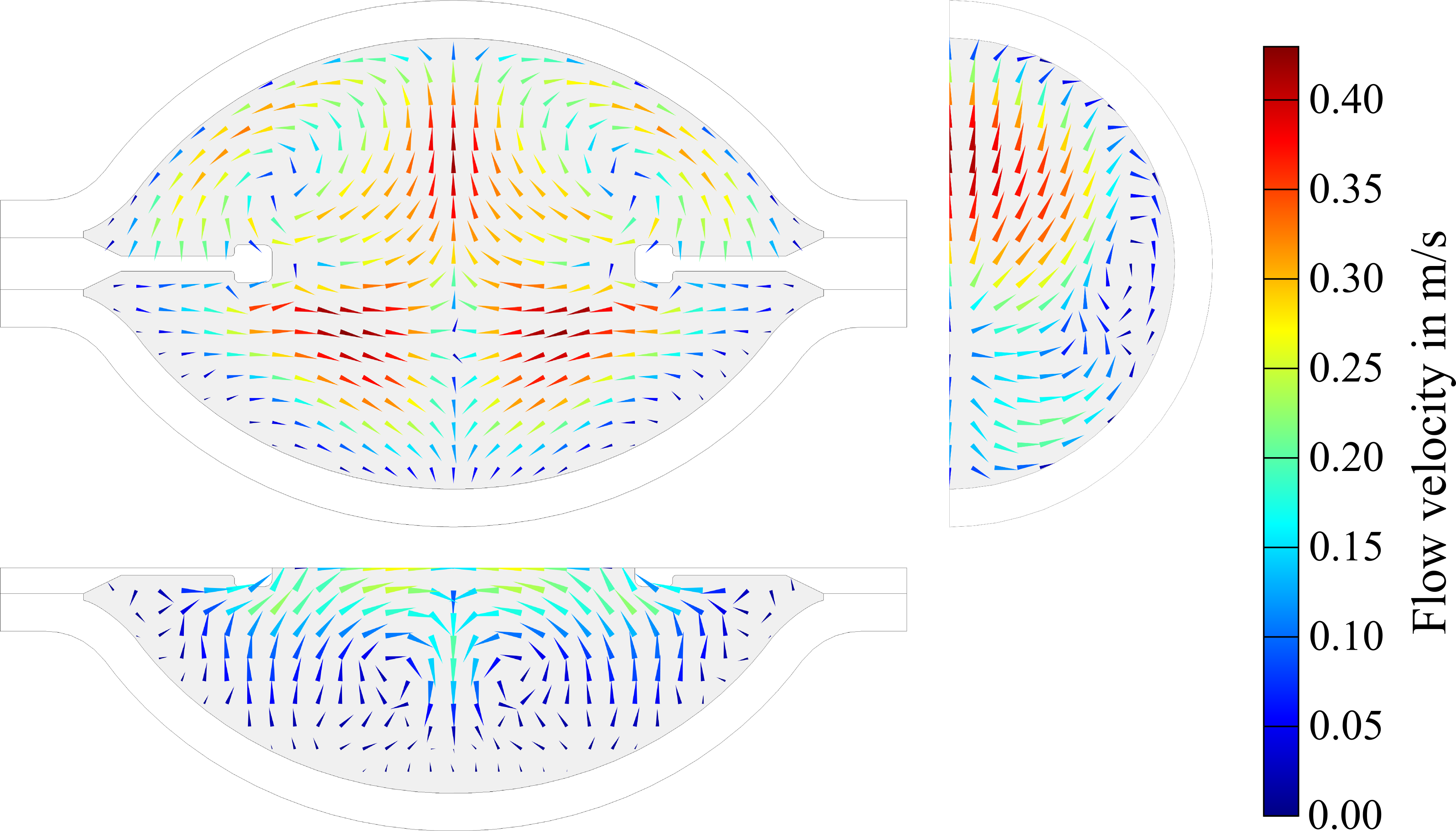}
\caption{Fluid flow patterns in three orthogonal planes through the arc tube's center. The upper diagram corresponds to the lower branch of the response curve ($f_{\rm e}=\ze{46.300}{kHz}$), the lower diagram to the upper branch ($f_{\rm e}=f_2=\ze{46.313}{kHz}$, jump frequency). The bottom view of both figures shows the horizontal plane ($x$-$y$-plane), whereas the top view shows the vertical planes (left: $y$-$z$-plane, right: $x$-$z$-plane). 
\label{fig:RecursionVelocity46300Hz_AllPlanes}}
\end{figure}

Figure \ref{fig:RecursionVelocity46300Hz_AllPlanes} shows two flow fields inside the arc tube. The upper set of images (side view, front view, top view) corresponds to the lower branch of the response curve at an excitation frequency just below the jump frequency ($f_2=\ze{46.313}{kHz}$). The absolute values of the velocity are rather small. The maximal velocity of $\ze{90}{mm/s}$ is located in the center of the arc tube, where the highest temperatures occur. The asymmetry in the horizontal plane is due to the $5\grad$ tilting. The pattern clearly has the characteristics of a buoyancy driven flow field. This is in contrast to the flow field depicted in the lower set of images, which corresponds to the upper branch of the response curve at the jump frequency $f_2$. Here, the flow in the vicinity of the negative $z$-axis points into the downward direction. This is typical for AS (see \cite{Dreeben.2008,0022-3727-48-25-255501}). The velocity of the flow is much higher and the asymmetry has vanished because the tilting does not affect AS and the AS force dominates over the buoyancy force.

\begin{figure}
\centering
\hspace*{0mm}
\begin{tabular}{lc}
\hline
\raisebox{-3mm}{\bf Lower branch} &   \\ 
\hspace*{5mm}\raisebox{8mm}{$\stackrel{\displaystyle f_{\rm e}=\ze{45.000}{kHz}}{\bar{T}=\ze{2828}{K}}$} & \includegraphics[width=0.45\linewidth]{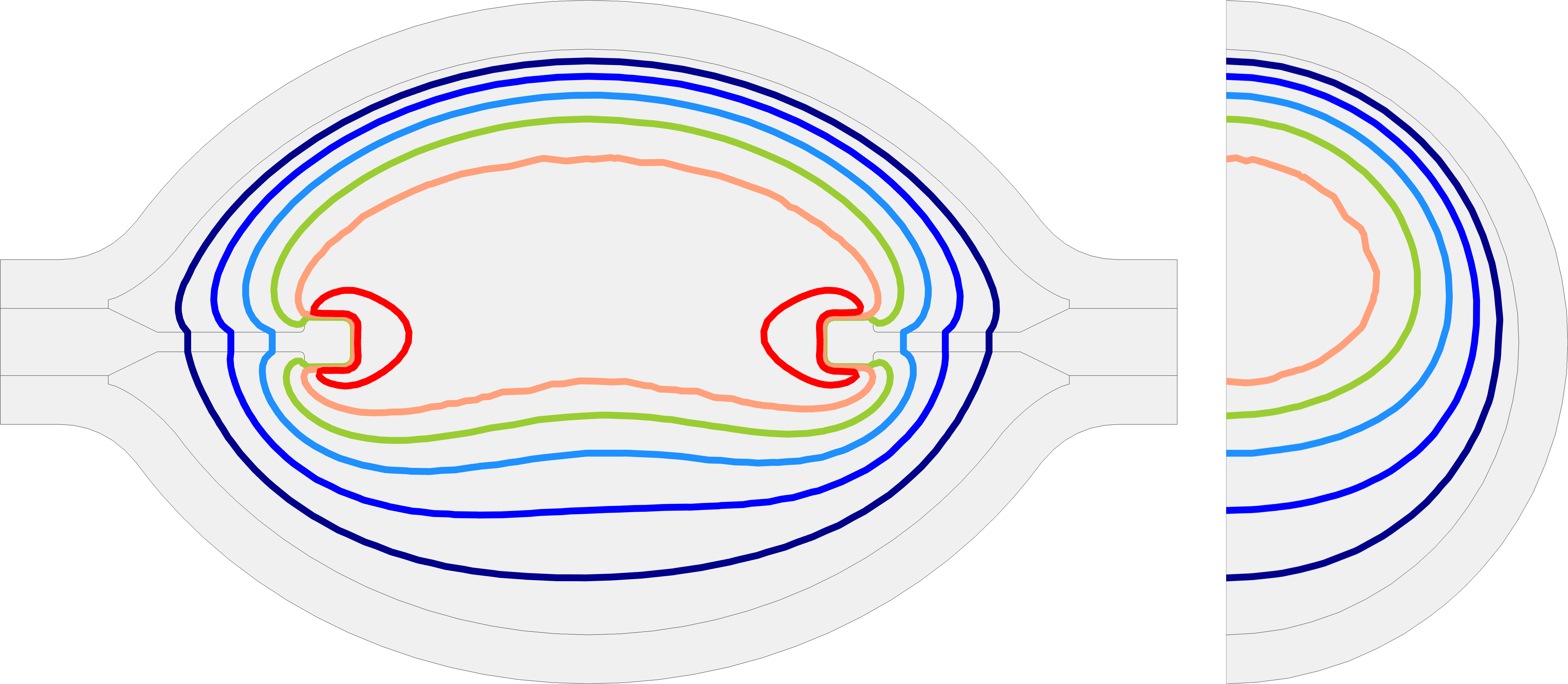}  \\ 
\hspace*{5mm}\raisebox{8mm}{$\stackrel{\displaystyle f_{\rm e}=\ze{46.700}{kHz}}{\bar{T}=\ze{2826}{K}}$}  &  \includegraphics[width=0.45\linewidth]{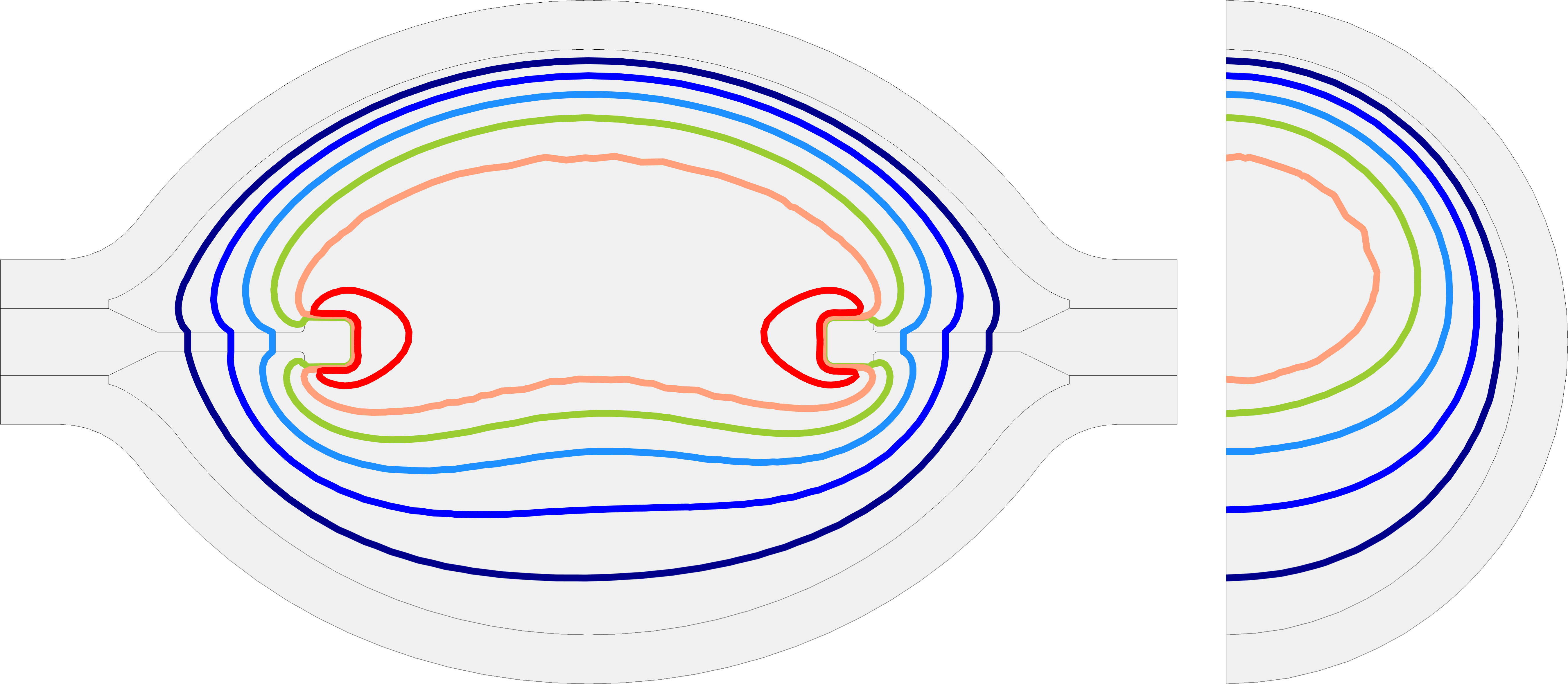}  \\ 
\hspace*{5mm}\raisebox{8mm}{$\stackrel{\displaystyle f_{\rm e}=\ze{47.141}{kHz}}{\bar{T}=\ze{2822}{K}}$}  & \includegraphics[width=0.45\linewidth]{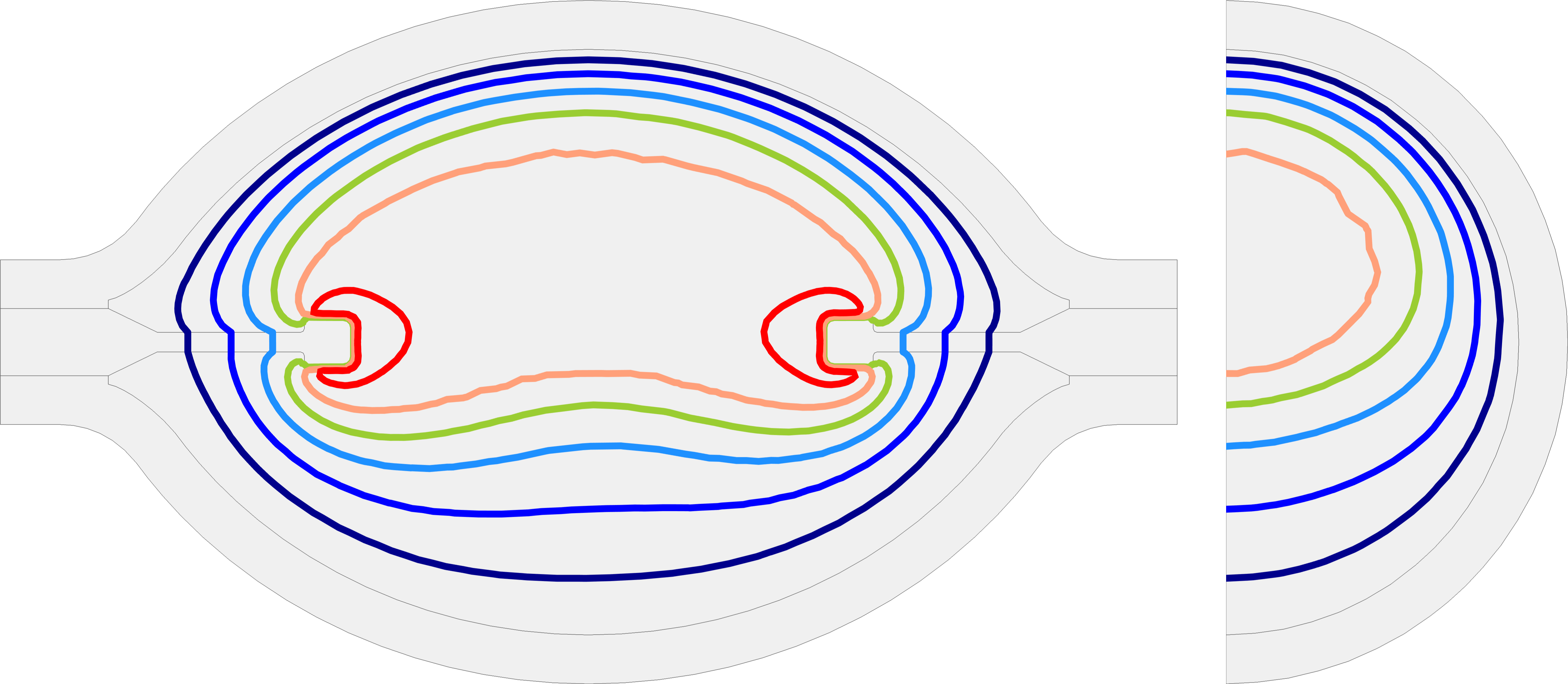}  \\ 
\hline
\raisebox{-3mm}{\bf Upper branch} &   \\ 
\hspace*{5mm}\raisebox{8mm}{$\stackrel{\displaystyle f_{\rm e}=\ze{46.313}{kHz}}{\bar{T}=\ze{2726}{K}}$} & \includegraphics[width=0.45\linewidth]{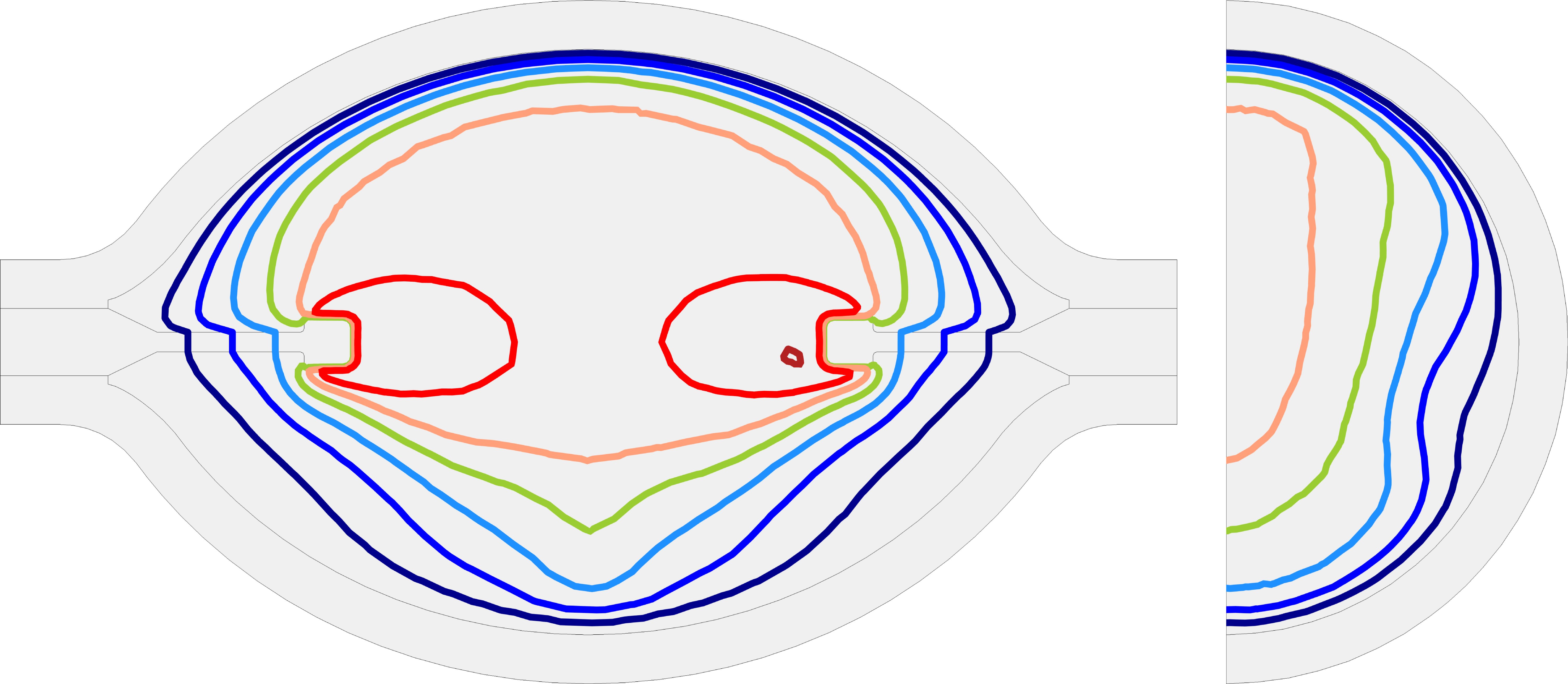}  \\ 
\hspace*{5mm}\raisebox{8mm}{$\stackrel{\displaystyle f_{\rm e}=\ze{46.700}{kHz}}{\bar{T}=\ze{2759}{K}}$} & \includegraphics[width=0.45\linewidth]{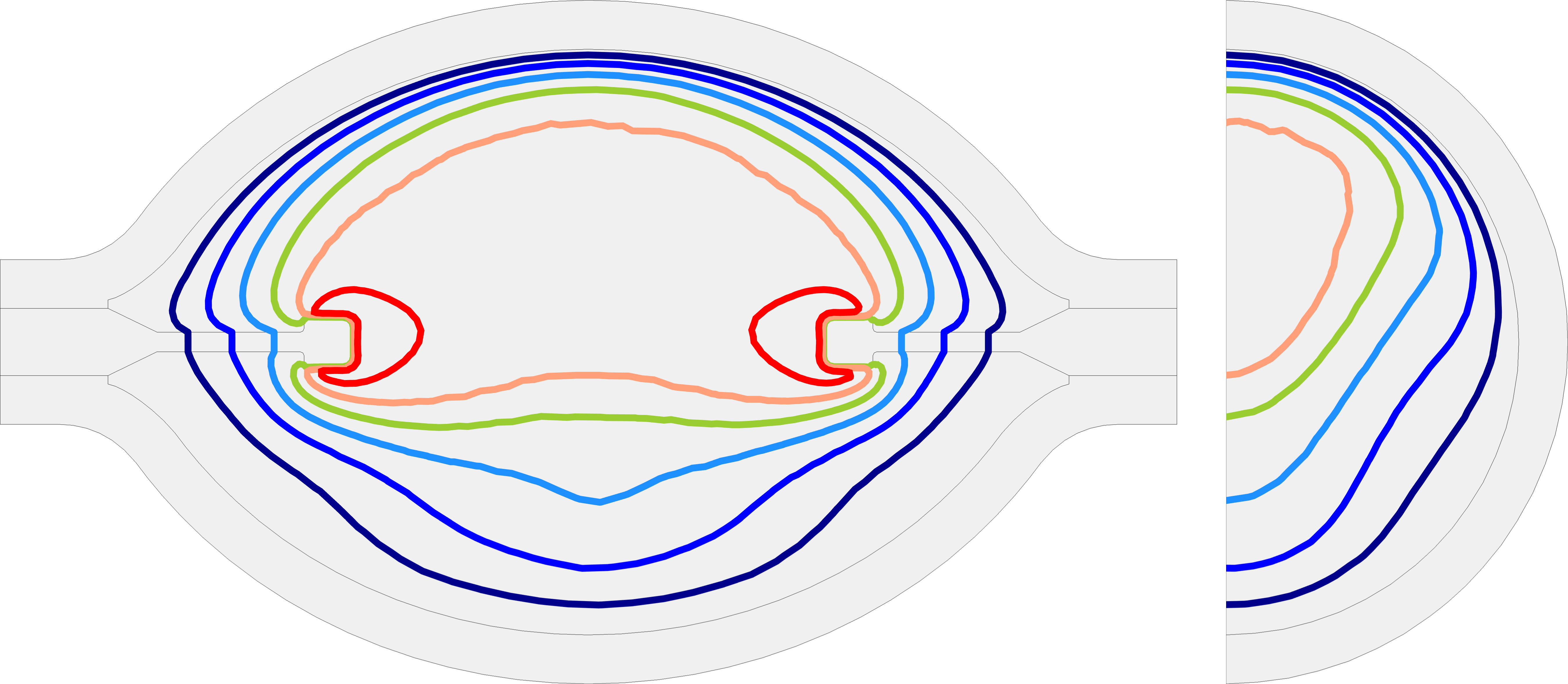} \\ 
\hspace*{5mm}\raisebox{8mm}{$\stackrel{\displaystyle f_{\rm e}=\ze{47.750}{kHz}}{\bar{T}=\ze{2817}{K}}$} & \includegraphics[width=0.45\linewidth]{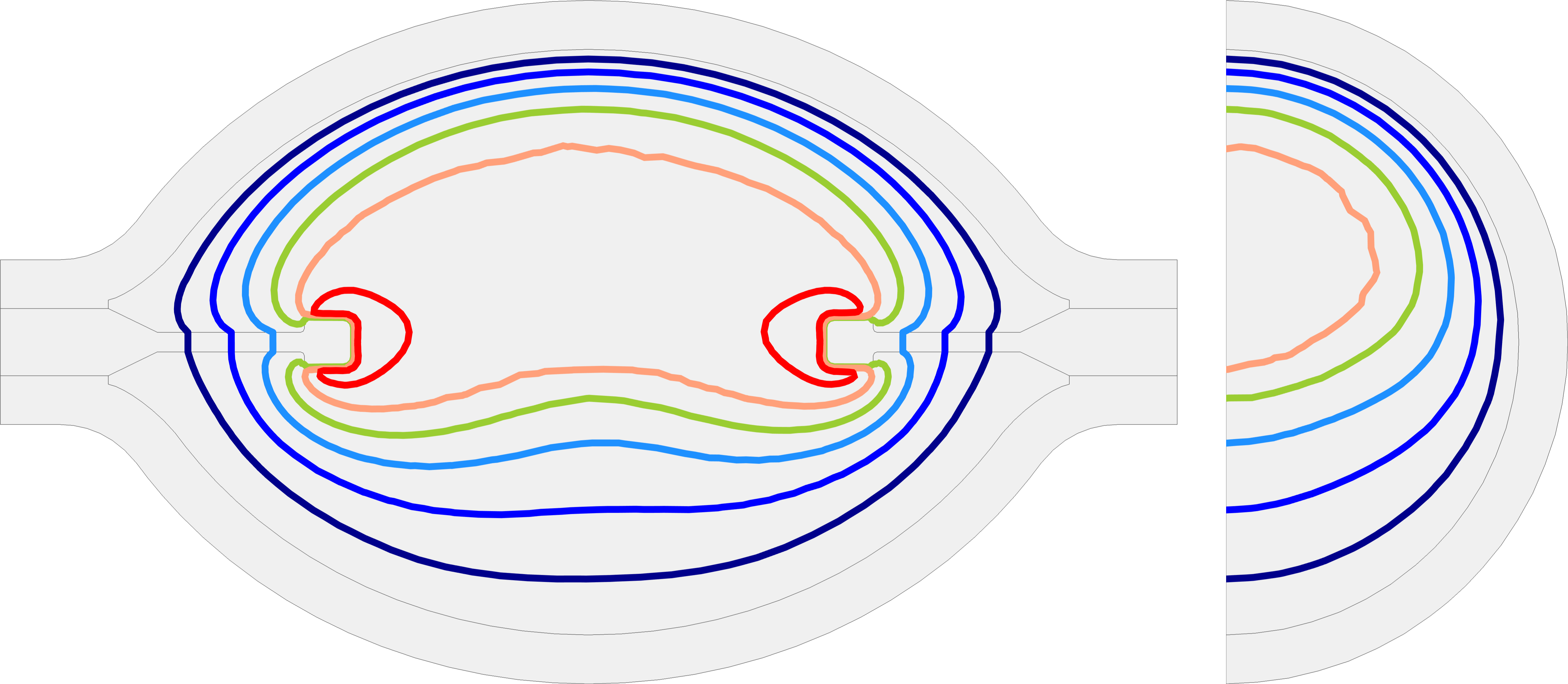}  \\
\hline
\end{tabular} 
\caption{Temperature profiles in the center planes of the arc tube. Top: Lower (ascending) branch (frequencies $\ze{45.000}{kHz}$ (below region of bistability), $\ze{46.700}{kHz}$  (inside region of bistability), $f_{\rm e}=f_1=\ze{47.141}{kHz}$ (high frequency margin of region of bistability)). Bottom: Upper (descending) branch (frequencies $f_{\rm e}=f_2=\ze{46.313}{kHz}$ (low frequency margin of region of bistability), $\ze{46.700}{kHz}$, $\ze{47.750}{kHz}$). The contour lines correspond to temperatures starting from $\ze{2000}{K}$ in $\ze{500}{K}$ steps. $\bar{T}$ denotes the mean value of the temperature taken over the interior of the arc tube.}
\label{fig:TemperatureProfiles1}
\end{figure}

How does the changed fluid flow alter the temperature profile inside the arc tube? In Figure \ref{fig:TemperatureProfiles1} various temperature profiles are depicted. The three profiles corresponding to the lower branch (ascending frequency) are virtually identical. For all frequencies belonging to the lower branch, whether below or inside the region of bistability, the temperature profiles look alike. This is different for the upper branch. The profile corresponding to a frequency well above the region of bistability looks similar to those corresponding to the lower branch. A close look reveals subtle differences: The temperature near the electrodes is lower than in the lower branch profiles. In addition, some contour lines are slightly deformed into downward direction. When the frequency decreases, theses deformations become very pronounced. Obviously, the AS flow field drags the high temperature plasma towards the lower part of the arc tube. In the upper part of the arc tube, where the velocity is very high and upward-directed, the hot plasma is pushed towards the wall. Here, the contour lines are very dense and, therefore, the gradient of the temperature is very large. The plasma arc is stretched in the vertical direction. At the same time it becomes narrower in its horizontal extension.

The large temperature gradient is connected to an increased heat loss \cite{Waymouth.1971b} that results in a temperature decrease. The mean values of the temperature inside the arc tube displayed in Figure \ref{fig:TemperatureProfiles1} confirm this effect. The temperature values in the lower branch, where the impact of AS is insignificant, are practically identical. In the upper branch almost the same average temperature has been calculated at $f_{\rm e}=\ze{47.750}{kHz}$, where AS is weak. The two remaining temperature values of the upper branch are essentially lower and, hence, confirm that the drop of the temperature is correlated with the strength of the AS effect.

In the high temperature region ($\ge\ze{3500}{K}$) the electrical conductivity of the arc tube content increases with the temperature \cite{0022-3727-48-25-255501,Flesch.2006}. Therefore, a temperature drop leads to an increase of the electric resistance $R$. Since the power $P$ in the simulations is kept fixed, the higher resistance is, according to $P=U^2/R$, connected to an increasing voltage $U$. This explains the similar behavior of the acoustic pressure and the voltage drop in Figures \ref{fig:PressVsXFEF} and \ref{fig:VoltvsXFSimExp}.

\subsection{Analogy to the Duffing Oscillator}
The drop of the temperature discussed in the previous section results in a decrease of the static pressure $P$. The pressure, in turn, is closely related to the modulus of compression $K$ \cite{Morse.1968}. 
To make a connection to the Duffing oscillator, we expand $K(P)$ in a power series around the nominal pressure value $P_0$:
\nfm{K(P)=K_1+K_2(P-P_0)+K_3(P-P_0)^2+...}
$K(P)$ and $K_1$ (corresponding to the linear spring constant $k_1$) have the dimension of a pressure and $K_3$ (corresponding to the cubic stiffness parameter $k_3$) has the dimension inverse to a pressure. To be compatible to the Duffing oscillator, we set $K_2=0$. 

The acoustic pressure $p$ corresponds to the displacement $y$ in the Duffing equation. We tentatively assume that the Equation (\ref{eq:Backbone1}) for the backbone curve can be translated from the oscillator problem to acoustics. Then we obtain
\nfm{p_{\rm P}(\Omega_{\rm P})=\sqrt{\frac{8K_1(\Omega_{\rm P}-1)}{3K_3}},\label{eq:Backbone2}}
where $p_{\rm P}$ stands for the peak acoustic pressure amplitude.
This assumption might be reasonable since the dimensions of Equation (\ref{eq:Backbone2}) match.

\begin{figure}
\centering
\includegraphics[width=0.9\linewidth,trim=0 0 0 0,clip]{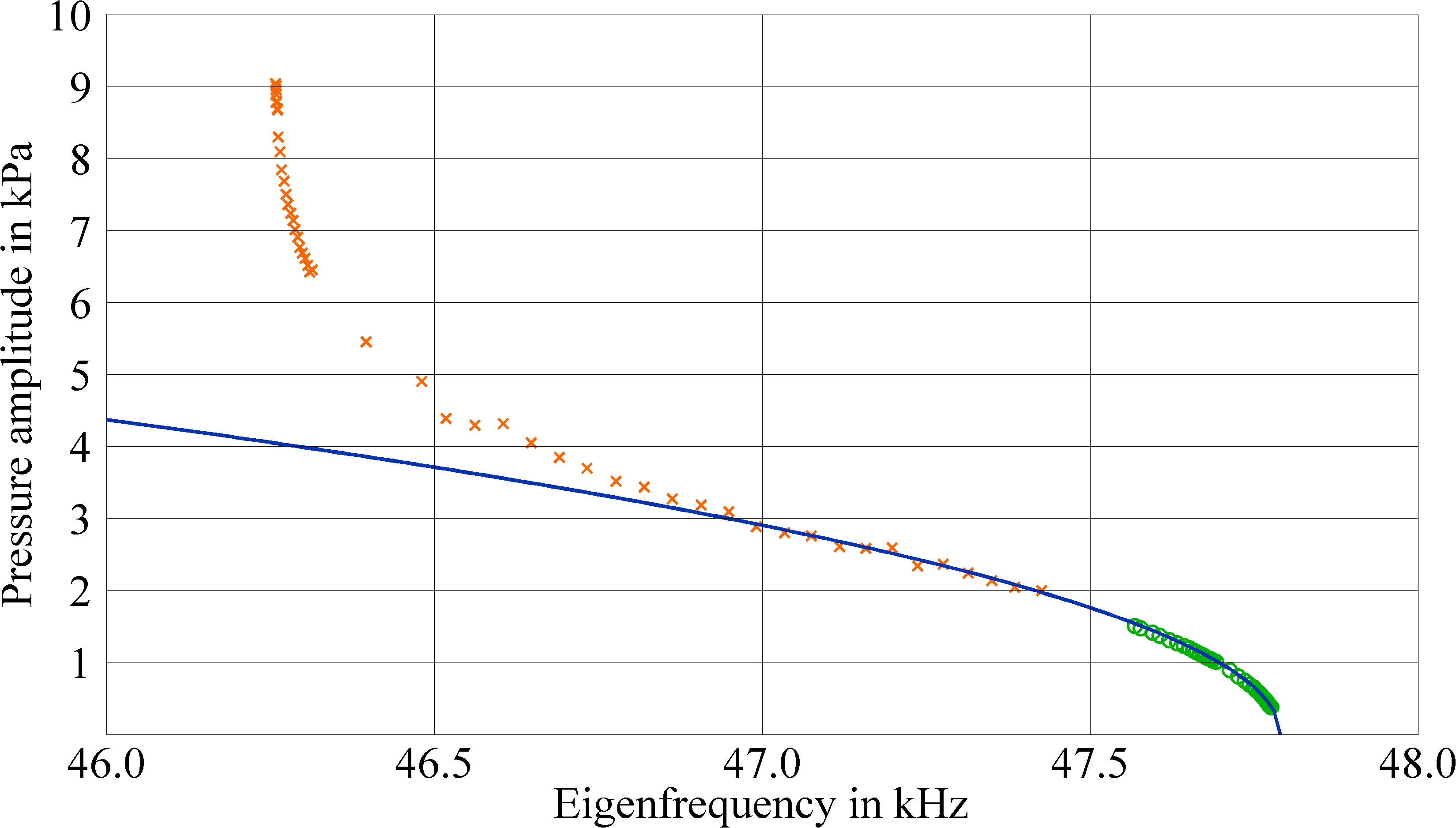}
\caption{Fit of Equation (\ref{eq:Backbone2}) to the backbone curve of Figure \ref{fig:PressVsXFEF}. The fit is based on the data depicted with green circles.
\label{fig:BackboneFit1}}
\end{figure}

In Figure \ref{fig:BackboneFit1} a fit of Equation (\ref{eq:Backbone2}) to the backbone curve of Figure \ref{fig:PressVsXFEF} is depicted. The two fit parameters are $\omega_0$ and the ratio $\frac{K_1}{K_3}$. The fit is based on the up-ramping data only, because we cannot expect the Duffing oscillator description to be valid near the eigenfrequency (see end of Section \ref{sec:AcousticResponse}). The fit is excellent and the results for the fit parameters yield $f_0=\omega_0/2\pi=\ze{47.79}{kHz}$ and $K_1/K_3=\ze{-191.3}{kPa^{2}}$. $K_1$ can be estimated from $c=\sqrt{\frac{K_1}{\rho_0}}$. For an eigenfrequency of about $\ze{48}{kHz}$ the averaged speed of sound roughly is $\ze{413.6}{m/s}$ and the density $\ze{24.89}{kg/m^3}$. This leads to $K_1\approx\ze{4258}{kPa}$ and $K_3\approx\ze{-22.26}{kPa^{-1}}$. The negative sign of $K_3$ is characteristic for the softening effect.

\section{Conclusions}
The behavior of HID lamps near an acoustical eigenfrequency has been examined by FE simulation and experiment. The numerical and the experimental results show a hysteresis well known from the Duffing oscillator with a softening spring: When ramping the excitation frequency upward/downward the voltage drop between the electrodes jumps to a higher/lower value at certain critical frequencies, and it seems plausible that the jumps are the cause of light flicker. The stiffness parameters have been estimated from a fit to the simulated backbone curve. The results from the model and the experiment are in good accordance, though the impact of AS is strongly underestimated by the simulations. The observed softening of the restoring force opens ways to cure the flicker problem of HID lamps: The literature describes measures that can be taken to counter the softening effect  \cite{Fahsi.2009,thomsen2002some,belhaq1999quasi}. We hope to solve the light flicker problem in HID lamps with one of these remedies.

\vspace{3mm}
{\noindent\bf Acknowledgment:} 
This research was supported by the German Federal Ministry of Education and Research (BMBF) under project reference 03FH025PX2 and Philips Lighting. We are indebted to Klaus Spohr and Thorsten Struckmann for discussions.

\pagebreak
\bibliographystyle{unsrt}
\bibliography{HIDLiteratur}   

\end{document}

%% file: Vorspann.tex
\usepackage{amsmath}
\usepackage{amsfonts}
\usepackage{amssymb}

\usepackage[latin1]{inputenc}
\usepackage{latexsym} 
\usepackage{graphicx}
\usepackage{psfrag}
\usepackage{exscale}
\usepackage[permil]{overpic}       
\usepackage[vflt]{floatflt}        
\usepackage{times}
\usepackage{makeidx}               
\usepackage{color}
\usepackage{framed,color}
\usepackage{wasysym}
\usepackage{rotating}

\makeindex   

   {\begin{list}{}{
      \settowidth{\labelwidth}{\textbf{#1}}
      \setlength{\leftmargin}{\labelwidth}
         \addtolength{\leftmargin}{\labelsep}
      \setlength{\parsep}{0.5ex plus0.2ex minus0.2ex}
      \setlength{\itemsep}{0.3ex}
      }}
   {\end{list}}

\newcommand{\ze}[2]{#1\;{\rm#2}} 

\newfont{\tensy}{cmsy10}           

\newcommand{\grad}{^\circ} 
                                        
\newcommand{\fm}[1] 
{\begin{displaymath}#1\end{displaymath}}

\newcommand{\nfm}[1] 
{\begin{equation}#1\end{equation}}

\newcommand{\nmzfm}[1] 
{\begin{eqnarray}#1\end{eqnarray}}

\newcommand{\mzfm}[1] 
{\begin{eqnarray*}#1\end{eqnarray*}}

\newcommand{\gfm}[1] 
{\begin{displaymath}\fbox{$\displaystyle #1$}\end{displaymath}}

\newcommand{\ngfm}[1] 
{\begin{equation}\fbox{$\displaystyle #1$}\end{equation}}

\newcommand{\mzgfm}[1] 
{\begin{displaymath}\fbox{$\begin{array}{rcl} #1\end{array}$}\end{displaymath}}

\newcommand{\figcapt}[2] 
{\renewcommand{\normalsize}{\small}\textbf{\caption[#1]{\textmd{#2}}}}

\newcommand{\fif}[1] 
{\mathbf{#1}}

\newcommand{\bi}[4] 
{\int_{#1}^{#2}{#3}\,{#4}}

\newcommand{\ubi}[2] 
{\int{#1}\,{#2}}

\definecolor{shadecolor}{gray}{.90} 

\newtheorem{beispiels}{Beispiel}[section] 

\newtheorem{aufgabeS}{Aufgabe}[section]

\newcommand{\qM} 
{\left[\ddots\right]}

\newcommand{\SV} 
{\left[\vdots\right]}